\documentclass[aps,pre]{revtex4}

\usepackage{graphicx}
\usepackage{amssymb,amsfonts,amsmath,color}

\begin{document}

\title{Stability of two-species communities:  drift, environmental stochasticity, storage effect and selection.}

\author{Matan Danino, David A. Kessler and Nadav M. Shnerb}

\affiliation{Department of Physics, Bar-Ilan University,
Ramat-Gan 52900, Israel.}

\begin{abstract}
\noindent The dynamics of two competing species in a finite size community is one of the most studied problems in population genetics and community ecology. Stochastic fluctuations lead, inevitably, to the extinction of one of the species, but the relevant timescale depends on the underlying dynamics. The persistence time of the community has been calculated for neutral models, where the only drive of the system is drift (demographic stochasticity) and for models with strong selection. Following recent analyses that stress the importance of environmental stochasticity in empirical systems, we present here a general theory of persistence time of two-species community where drift, environmental variations and time independent selective advantage are all taken into account. \\


\end{abstract}

\maketitle

\section{Introduction}

One of the main contemporary challenges of the life sciences is to understand the factors that allow for the maintenance of biodiversity \cite{sachs2009biodiversity,chesson2000mechanisms}. A fundamental proposition in population genetics and community ecology, the competitive exclusion principle~\cite{hutchinson1961paradox,stomp2011large}, suggests that when two genetic alleles or two biological species compete for the same resources only one  species/allele  will survive. Despite its theoretical importance and its firm mathematical foundations, many natural systems appear to violate this principle, allowing for coexistence of many competing species or (higher than expected) polygenic variations.

In community ecology, the simplest explanation for such a situation is niche partitioning, meaning that multiple limiting resources may give rise to a collection of species, ranging up to  the number of resources, with each species excelling with respect to one resource~\cite{tilman1982resource}.  However the identification of limiting resources is difficult in practice, and in some cases (tropical trees~\cite{ter2013hyperdominance}, fresh-water plankton~\cite{hutchinson1961paradox,stomp2011large} and coral reef~\cite{connolly2014commonness}) the niche-partitioning mechanism seems implausible. An understanding of possible alternative coexistence-promoting mechanism is a subject of much interest both in community ecology \cite{chesson2000mechanisms} and population genetics.

Taking into account the inherent stochasticity in biological populations dynamics, one realizes that the biodiversity puzzle is, in fact, a question about \emph{time scales}. The dynamics of every population admits an absorbing state: once the species goes extinct, it cannot recover again. Accordingly, every biosystem suffers from a continuous loss of life forms, a process that reduces its diversity. Biodiversity equilibrates  when the rate of extinction matches the rate at which new types are introduced into the system as a result of speciation or mutation events (or, for a local community, migration from a regional pool).

 An important theoretical framework in which this insight is implemented is the neutral model, both in its well-mixed form \cite{kimura1985neutral} (in genetics) and in its spatial, mainland-island version \cite{hubbell_book,maritan1,TREE2011} (in community ecology). The neutral model assumes that all species are demographically equivalent (no selective advantage) and that species abundance varies only due to genetic/ecological drift (demographic stochasticity). When two species compete under these conditions, the persistence time of the community (the time until one of them goes extinct, also known as the absorption time \cite{ewens2012mathematical}) is, on average, $N$ generations, where $N$ is the size (number of individuals) of the community. If the timescale on which new types are introduced into the system (by speciation, mutation or migration) is comparable to the persistence time, the typical number of species will be larger than one.

Recently, a series of studies showed that the abundance variations in empirical ecological communities are much stronger than the predictions of the neutral model \cite{kalyuzhny2014niche,kalyuzhny2014temporal,chisholm2014temporal}. This appears to reflect the presence of \emph{environmental} stochasticity,  i.e.,  the random variations in species relative fitness caused by fluctuations in exogenous factors like precipitation, temperature, predation pressure and so on \cite{lande2003stochastic}. Demographic noise accounts for the stochastic factors that affect each individual independently, so the average fitness of a population is fixed and abundance fluctuations per generation scale like the square root of population size. Conversely, under environmental stochasticity the fitness of all individuals in a certain population fluctuates in a correlated manner, leading to much stronger abundance variations.

Based on this observation, a time-averaged neutral theory of biodiversity has been suggested \cite{kalyuzhny2015neutral,danino2016effect}, where environmental stochasticity affects the system but species are still symmetric since each species' fitness, \emph{when averaged over time}, is identical. Again, species go extinct at a certain rate, now determined by both demographic and environmental stochasticity, and biodiversity reflects the balance between extinction and speciation/migration rates. This model has been shown to fit quite nicely the static and dynamic characteristics of a (local) community of  tropical trees \cite{kalyuzhny2015neutral,fung2016reproducing}; both the species abundance distribution and the abundance variations are similar to the predictions of the model as obtained from numerical simulations.

The inclusion of environmental stochasticity into the model opens up, again, the timescale problem. Environmental stochasticity is stronger than drift, and overshadows its effect in large abundance populations. Environmental noise at first sight appears to be a  destabilizing factor: it increases the rate in which the system jumps from one state to another, and hence shortens the time until a population reaches a low-abundance state and goes extinct. The naive expectation, thus, is that environmental stochasticity shortens the persistence time, though one would like to quantify this argument and to find the $N$ dependence explicitly.

However, under some circumstances environmental stochasticity may become a \emph{stabilizing} mechanism, as suggested by Chesson and collaborators \cite{chesson1981environmental,hatfield1989diffusion,hatfield1997multispecies}.
These authors show that environmental variations may enhance the chance of invasion of low-abundance species via the \emph{storage effect}: rare species, when compared with common species, have fewer per-capita losses when their fitness is low and more gains when their fitness is high. As a result, the system may support a stable equilibrium: every species' abundance fluctuates, but all are peaked about some finite value by a noise-induced stabilizing force.

Chesson and coworkers introduces the \emph{lottery model}, a minimal model that captures the essence of the storage effect, and analyzed its stability properties. However, they considered a system with pure environmental noise and without demographic stochasticity. In such a system there is no extinction per se, as population density may take arbitrarily small values. Accordingly, the criteria they used to define a stable equilibrium was the normalizability of the probability density function. This strategy did not allow them to calculate extinction times, making it impossible to analyze diversification rates.

In a recent paper \cite{hidalgo2016species}, Hidalgo et. al. considered the two-species community persistence problem in the presence of environmental stochasticity, with and without storage. Like  \cite{chesson1981environmental}, they analyzed the dynamics of a two-species community with pure environmental noise, such that the number of individuals belonging to each of the species is not necessarily an integer. Extinction was defined, in their work, as the case where the density of a species goes below one individual; for a community of $N$ individuals extinction happens when the density is smaller than $1/N$. Looking at the system under  dichotomous (telegraphic) environmental noise, they were able to calculate the large $N$ asymptotic behavior of extinction times for a time-averaged-neutral community. This procedure may be justified, as in the large $N$ limit the discreteness of agents is less important, since the number of individuals is typically large, so one may replace the demographic noise by a threshold value at $1/N$. To close the gap between the asymptotic behavior at large $N$ and the regimes where demographic noise is important, Hidalgo and coworkers suggested the existence of (one or two parameter) scaling functions and provide a numerical evidence to support their conjecture.

Here we solve the persistence time problem in all its glory, taking into account explicitly both demographic and environmental stochasticity. This allows us to extend the theory suggested by Hidalgo et al. in the following senses:

\begin{enumerate}
  \item An explicit, closed form for the scaling functions (in terms of a single or a double integral) is derived, so the answer covers all the range of parameters. In particular our formulas converge to the pure demographic limit when the environmental stochasticity vanishes.
  \item The expressions suggested in \cite{hidalgo2016species} for the large $N$ limit are recovered, but we can calculate also subleading terms in this asymptotic series. This allows us to identify the parameter region where the asymptotic is accurate, and to suggest simple analytic approximations that cover  a much wider region of parameters.

  \item We can calculate the persistence time for a single mutant. This is an important quantity, as it sets the threshold for clonal interference and may be relevant to the small island effect in island biogeography (see next section).
\end{enumerate}

Moreover, we have  extended the work of \cite{hidalgo2016species}  to include the case where one species has a time independent selective advantage with respect to the other species, superimposed on the environmental variations.

This paper is organized as follows. In the next section we provide a few basic intuitive arguments and a summary of the main results. Section \ref{demo} deals with the case of pure demographic noise, in section \ref{maritan} we consider the case of demographic and environmental noise without storage effect, and in \ref{chesson} the case with storage effect. Section \ref{selection} is devoted to the effect of selection on persistence time when it acts against the storage mechanism, and is followed by a discussion section. For the sake of completeness we describe in appendices  the results for a system with selection and pure demographic noise (Appendix \ref{apb}) and selection with both environmental and demographic noise, but without storage (Appendix \ref{apc}).

\section{Intuitive arguments, glossary and summary of the main results}

In this section we explain the main issues considered along this paper, introduce the notations, provide a glossary (see Table 1) and briefly sketch the main results.

\subsection{Persistence time: definition and the importance of initial conditions}

Throughout this paper we consider a two ``species" (genetic types, zoological species, bacterial strains) playing a zero-sum game in a finite size community: one individual is chosen at random to die and is removed, with its slot being captured by an offspring of another individual, chosen with some probability that  reflects its (momentarily) relative fitness. Such a stochastic birth-death process leads, inevitably, to the extinction of one of the species, and our goal is to calculate how much time, on average, it takes for the system to reach this monotypic state (the time to absorption), given the size of the community, $N$, and the initial conditions, in which species 1, say, is represented by $n$ individuals and species 2 has $N-n$ individuals.

We do not calculate here the fixation time, which is defined as the average time it takes to a specific species to reach $N$ \emph{conditioned on} its  success. The quantity we are looking at measures the persistence time of a community (or the time to fixation or loss).

While we provide expressions for any initial condition, two special cases are of particular importance. One is the case of a single mutant introduced into a system, i.e., $n = 1,  \ N-n = N-1$. This is the relevant case when a new type appears in a single mutation/speciation event, or when the system is subject to weak immigration from a regional pool. The biodiversity (species richness, genetic polymorphism) of such a system depends on the balance between the rate of appearance of new types (rate of mutation/migration/speciation) and the average time it takes for a single mutant to reach either zero or $N$. If the rate of mutation is small (in units of inverse persistence time), the community will be monotypic most of the time. If the rate is high, the typical state has more than one species, and thus the ratio between these two rates determines the threshold for clonal interference \cite{gerrish1998fate}. By the same token, a recently proposed explanation for what known in island biogeography as  the small island effect suggests that the number of niches per island is fixed and for small islands the rate of absorption is smaller that the rate of immigration \cite{chisholm2016maintenance}. Under these assumption, an island is ``small" as long as the immigration rate is smaller than the inverse of the persistence time calculated here.

Another interesting scenario is the case where both $n$ and $N-n$ are initially large. This may happen when the community was subject to strong invasion, when speciation occurs allopatrically and then the two groups mix again, for protracted speciation \cite{rosindell2010protracted} or in experiments in which two bacterial strains or two vegetation species are mixed \cite{jiang2007temperature}. One may get insight into the persistence time in this case by  identifying the  initial state for which this time is maximized (e.g., in a symmetric system this will be the initial state $n = N/2$) and calculating the persistence time for this state. In what follows we provide simple expressions for these two cases: the maximum value of the expected persistence time and the expected persistence time for a single mutant.

\subsection{Demographic stocahsticity (drift) and the neutral model}

The reproductive success of any individual is subject to many random events. In our system, this property is reflected in the random nature of the birth-death process. When the two species in the community are demographically identical, and population variations are driven solely by demographic noise (one individual is chosen to die and another is chosen to reproduce, both choices are random and the species affiliation does not play any role), the dynamics is known as  \emph{neutral}. Under neutral dynamics the maximum value of the expected persistence time (measured in units of generations, see below) is known to be proportional to $N$ \cite{crow1970introduction}. We review the features of the neutral model in section \ref{demo} below.

\subsection{Selection}

When one species has a selective advantage with respect to the other, its chance to capture the whole community is, of course, larger. In the game considered here we always kill one individual at random, with no distinction between species, but if species 1, say, has a selective advantage, the per capita chance of an individual of species 1  to capture the empty slot is larger than the chance of a species 2 individual. The difference between these two per capita chances is proportional to the selection parameter $\eta_0$.

On average, a species with selective advantage grows exponentially, so the population of the fitter species (say, species 1) will grow like  $n = exp(\eta_0 t)$, and the time it takes to  reach $N$ from $n = 1$ will scale like $\ln(N)/\eta_0$ \cite{crow1970introduction}. This is indeed the case, and the details of the calculation are given below (Appendix \ref{apb}).

\subsection{Environmental stochasticity}

The paper is focused on the effect of environmental stochasticity on the persistence time.  As the environmental conditions vary, the relative fitness of species may change.  Accordingly, for large populations the effect of environmental stochasticity is more important than the drift \cite{lande2003stochastic}. However, without demographic stochasticity there is no extinction; to calculate extinction times one must either consider explicitly the demographic noise, as we are doing here, or impose a threshold on the density at the level of one individual, as done in \cite{hidalgo2016species}.

When both species have the same average fitness, but the fitnesses fluctuate randomly so that  at any single moment one of them is superior (i.e., species fitness varies in time, but when averaged over time the two species are still symmetric) we speak about a system under environmental stochasticity and characterize it by two quantities: the squared amplitude of fitness variations $\gamma^2$ and the correlation time of the environment (in units of generations) $\delta$. Environmental stochasticity may be superimposed on a time independent fitness advantage of one species, and in that case we make a distinction between selection (the time independent component of the relative fitness) and stochasticity (fitness fluctuations with zero average over time).

The simplest way to think about a system driven by environmental stochasticity (with no selection) is to consider it as a random walk in the logarithmic space. When the environment gives an instantaneous fitness advantage to one species its population grows exponentially, and when the environment switches it shrinks exponentially, so overall the log-population is doing a random walk and the step size is proportional to $\gamma^2 \delta$. Since the typical time needed for a random walker to cross a distance $L$ scales like $L^2$, one expects that the maximum persistence time for a community of size $N$ will scale like $\ln^2(N)$ for large $N$. This asymptotic behavior was found by Hidalgo et. al \cite{hidalgo2016species}. Here, in section \ref{maritan}, we provide the expression for any $N$ and any initial conditions (Eqs. \ref{eq10} and \ref{eq11}),  calculate the leading corrections to the large $N$ result and identify the region in which this log squared asymptotics  is valid (Eqs. \ref{eq12} and \ref{eq12a}). We also show that the result for a single mutant initial condition has a different scaling (Eq. \ref{eq11c}) and explain why this happens.

\subsection{Noise induced stabilization: the storage effect}

Environmental stochasticity thus seems to be a destabilizing factor, as it increases the amplitude of abundance fluctuations and shortens the persistence timescale from $N$ generation (with pure demographic noise) to $\ln^2(N)$. However, under some circumstances the environmental stochasticity  stabilizes the system and increase substantially the persistence time. This possibility was discovered by Chesson and coworkers\cite{chesson1981environmental,chesson1994multispecies}, and is known as the storage effect.

The system considered here may provide a realization of the simplest example given for the storage effect, the Chesson-Warner lottery game. To do that, one may think about individuals as trees, say, and assume that the seed bank in the soil reflects the abundance of each species. Upon death of an individual one of the seeds is chosen to capture its location as an adult tree with a chance proportional to its fitness, so the overall chance of a species to increase its population reflects both its abundance and its instantaneous fitness, superimposed nonlinearly. As an example let us think about a "winner takes all" case, where the species with higher fitness wins the empty slot for certainty. Starting with 8 red individuals and 2 green in a community of 10, the chance to end up with 7 red and three green is 0.4 (a red is killed with probability 0.8, the green captures the slot with probability 1/2 since it is preferred half of the time) while the chance to end up in the 9:1 state is only 1/10. Rare species have a larger chance to grow in abundance just because they are rare.

On the other hand, if the dynamics of the system takes place in a series of duels, as described in the last subsection, the chance for an interspecific duel will be $32/90$, and so the probabilities to end up at 9:1 or at 7:3 will be the same: $16/90$, meaning that there is no preference to rare or common species and no storage effect.

The persistence time for a system with environmental noise and storage effect was considered by \cite{hidalgo2016species}, who suggested that for large $N$ it scales like $N^{1/\delta}$. Our analysis provides an expression for any $N$ and any initial condition (Eq. \ref{eq23}),  with an analysis of the  large $N$ behavior and the crossover to the $N^{1/\delta}$ regime (Eq. \ref{eq27}). We also consider  the same problem for a single mutant (Eq. \ref{eq27a}), showing that in this case the scaling of the persistence time with $N$ is also $N^{1/\delta}$, and that the difference between the maximum persistence time and the single mutant case is only a factor of $\gamma^2/2$.

\subsection{The interplay between stochasticity and selection}

The situations described so far correspond to the standard neutral model (perfect demographic equivalence, zero fitness differences between species and individuals) and to the time-averaged neutral model (TNTB) \cite{kalyuzhny2015neutral,danino2016effect}, where the species have the same fitness on average, but at each moment one species has higher fitness. In both cases, stochasticity (demographic or environmental) is the only driver of abundance variations.

However,  the generic scenario appears to be the case where a time independent selective advantage to one of the species is superimposed on environmental variations that may give instantaneous superiority to its opponent.

We consider this situation, with and without storage effect, in Section \ref{selection} and appendix \ref{apc}. For the case without storage we provide the general formula in (\ref{eq64}), but in the large $N$ limit selection alone determines the persistence time (Eq. \ref{eq70}). With storage the situation is much more interesting:  there is a scaled selection parameter $\tilde{s} = 2\eta_0/\gamma^2$, and as long as this parameter is smaller than one, noise induced stabilization wins and the time to extinction grows superlinearly with $N$ (like $N^{(1-\tilde{s})/\delta}$). These results are summarized in Eqs. (\ref{eq43}) (general formula) and (\ref{eq47}) (large $N$ asymptotics).

\begin{table}
\begin{center}
\caption {Glossary} \label{table1}
    \begin{tabular}{ | l |  p{10cm} |}

    \hline
    Term  &  Description \\ \hline
    $N$ &  number of individuals in the community (both species). \\ \hline
    $n$ &  number of individuals belonging to species 1 ($N-n$ belong to species 2). \\ \hline
    $x$ &  fraction of species 1, $x=n/N$.  ($1-x$ is the fraction of species 2)\\  \hline
    $\tau$ & correlation time of the environment, measured in units of elementary steps. \\ \hline
    $\delta$ & correlation time of the environment, measured in  generations.\\
    \hline
    $T(x)$ & mean persistence time for a two species system at $x$, when the environment is fixed in time. \\ \hline
    $S(x)$ & mean persistence time for a two species system at $x$, when the environment is fluctuating between two states, and the average is taken over both initial conditions and histories. \\ \hline
     $\eta_{i,j}$ & the relative fitness of species $i$ and $j$. $\eta_{2,1} = \eta_0  \pm \gamma$.  \\ \hline
     $\eta_0$ & the time-independent component of the fitness.   \\ \hline
     $\gamma$ & the amplitude of the fluctuating component of the fitness.  \\ \hline
    $G \equiv N \delta \gamma^2/2$ & scaled environmental stochasticity. \\ \hline
    $\tilde{s} \equiv 2\eta_0/\gamma^2 $& scaled selection. \\ \hline
    \end{tabular}

\end{center}

\end{table}

\section{Neutral dynamics with pure demographic stochasticity (drift)} \label{demo}

 To set the framework for the next sections and to clarify a few technical points, let us begin by considering the already studied \cite{crow1970introduction} case of a two species system with fixed community size $N$ and pure demographic noise. Species  $\# 1$ is represented by $n$  individuals and species $\# 2$ by $N-n$ individuals. During each elementary step one individual (chosen at random) dies (removed from the community, leaving a gap), and an offspring of another (again, randomly chosen) individual, is recruited into this gap. For such a fixed size community, with fixed death rate, the duration of each elementary birth-death event is $1/N$ so that the unit time is a generation.

If the birth-death event involves two conspecific individuals (one dies and the offspring of the other captures the vacant gap) the frequencies of the two species remain fixed. The chance of an interspecific ``interaction" (elementary event in which the frequencies of the two species is modified) is,
\begin{equation}
F(n) =F(N-n) = \frac{2n(N-n)}{N(N-1)} \approx 2x(1-x),
\end{equation}
where $x = n/N$ and $N \gg 1$ such that $N-1 \approx N$.

Without loss of generality, let us focus on the first species, represented at $t=0$, say, by $n_0$ individuals. Eventually the system must reach one of the two absorbing states, $n=0$ or $n=N$. The mean persistence time given the initial state, $T(n_0)$ depends on $n_0$ and satisfies the backward Kolomogorov equation (BKE) \cite{redner2001guide},
\begin{equation} \label{eq2}
T(n_0) = [ 1-F(n_0)]  \left( T(n_0)+\frac{1}{N} \right) + F(n_0) \left( \frac{1}{2}   T(n_0-1)+\frac{1}{2}   T(n_0+1) + \frac{1}{N} \right),
\end{equation}
with the boundary conditions
\begin{equation} \label{eq2a}
T(0)=T(N)=0.
 \end{equation}
 From now on we refer to $n_0$ simply as $n$, since Eq.  (\ref{eq2}) does not involve explicitly the dynamics of the system.

\begin{figure}
\includegraphics[width=16cm]{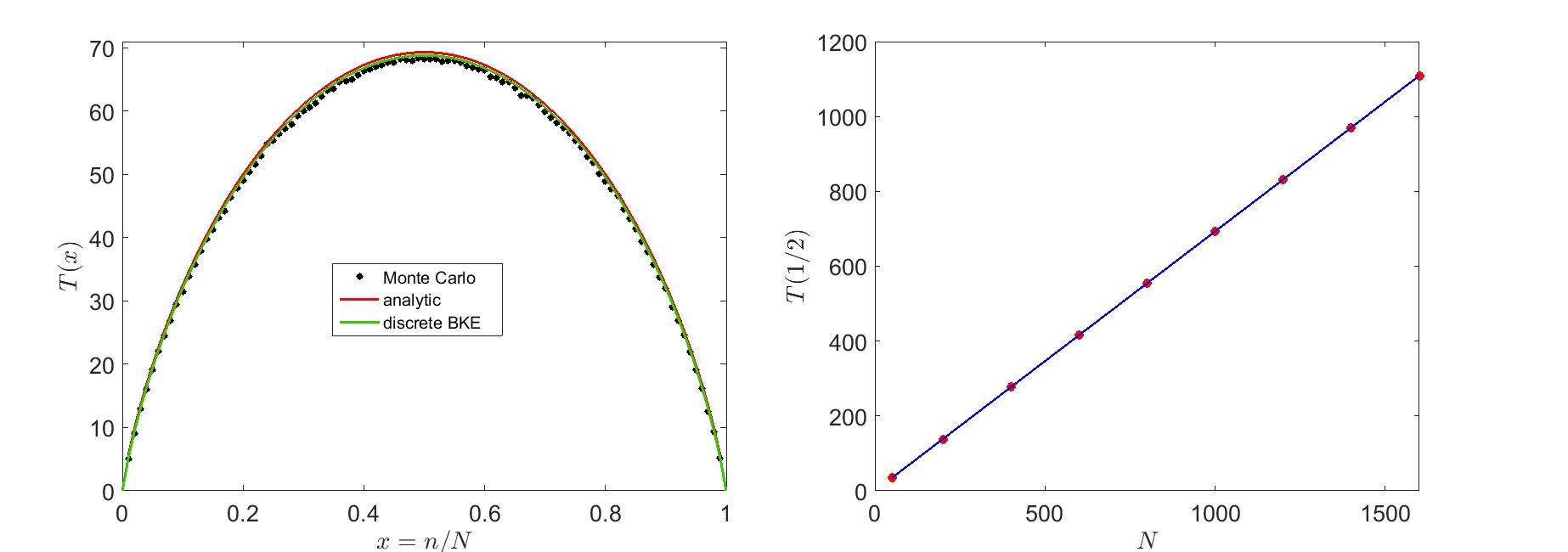}
\caption{A comparison between the results of a Monte-Carlo simulations, numerical solution of the backward Kolomogorov equation (see methods in Appendix \ref{methods}) and the analytic results (\ref{eq5}) for $T(x)$. In the left panel $T(x)$ is plotted against $x$ for a system with $N=100$, where the MC simulation reflect an average over $10000$ runs. In the right panel the maximum lifetime $T(1/2)$ is plotted as a function of $N$ (red filled circles) and is compared with the $T = N \ln(2)$ relationships predicted from (\ref{eq5}) for $x=1/2$. In both cases the agreement is perfect.}\label{fig1}
\end{figure}

 To write the BKE (\ref{eq2}) as a differential equation, we consider $n$ as a continuous variable and expand $T(n \pm 1)$ to  second order, obtaining
 \begin{equation} \label{eq3}
 -\frac{1}{N} = F(n) \cdot \left[ -T(n)+\frac{1}{2}   T(n-1)+\frac{1}{2}   T(n+1)  \right] \approx F(n) \cdot \frac{1}{2}\frac{\partial^2 T(n)}{\partial n^2}.
\end{equation}
Writing $x \equiv n/N$, this equation takes the form,
 \begin{equation} \label{eq4}
 -\frac{N}{x(1-x)} = \frac{\partial^2 T(x)}{\partial x^2}.
\end{equation}
Integrating twice and plugging in the boundary conditions (\ref{eq2a}) one finds,
 \begin{equation} \label{eq5}
T(x) = -N [x \ln(x) + (1-x) \ln(1-x)],
\end{equation}
meaning that the persistence time, for every fixed $x$, grows  linearly with $N$. Figure \ref{fig1} demonstrates the validity of Eq. (\ref{eq5}) when tested against two types  of numerical results: a numerical solution of Eq. (\ref{eq2}), written as $T_n =1/N + \sum_m {\cal M}_{mn} T_m$ and solved by matrix inversion, and a direct Monte-Carlo (MC) simulation of the underlying neutral process, averaged over many realizations. The details of these numerical techniques are given in Appendix \ref{methods}.

Two important features of (\ref{eq5}) are the maximum  persistence time obtained at $x=1/2$,
\begin{equation}
T(1/2) = N \ln(2),
 \end{equation}
and the persistence time of a single individual,
\begin{equation}
 T(1/N) = \ln(N).
 \end{equation}
If the per-birth chance of a single individual of a new type to enter the system  (for example, the chance of a newborn individual to be a mutant, or the originator of a new species, or the chance that an immigrant from a regional pool, belonging to a different type, replaces a dead individual) is $\nu$, then the typical number of species in the community will be unity as long as $\nu N \ln(N) <1$ (so that when a new type appears, it typically goes extinct or takes over the whole community before the next speciation/migration event), and the system typically will be found  in a state with more than one species if $\nu N \ln(N) >1$. The value $\nu_c = 1/[N \ln(N)]$ is thus the scale for \emph{clonal interference} \cite{gerrish1998fate,park2007clonal} and for the small island effect \cite{chisholm2016maintenance}.

In fact, one can solve the difference equation (\ref{eq2}) exactly by writing it as a first order difference equation for $W(n) \equiv T(n) - T(n-1)$,
\begin{equation}
-\frac{2N}{n(N-n)} = W(n+1)-W(n),
\end{equation}
so $$W(n) = W(1) - 2\sum_{k=1}^{n-1} \left( \frac{1}{k} + \frac{1}{N-k} \right).$$ Deriving $T(n)$ in the same way again applying the boundary condition $T(0) = 0$ and $T(N)=0$ one finds,
$$ T(n) = N H_n(N) - n H_n(n) - (N-n) H_n(N-n),$$ where $H_n \equiv \sum_{k=1}^n k^{-1}$ is the harmonic number. This expression converges to Eq. (\ref{eq5}) in the large $N$ limit. The maximum persistence time is,
\begin{equation} \label{eq5b}
T(N/2) = N\left( ln(2)-\frac{1}{2N} + {\cal O} \left[\frac{1}{N^2}\right] \right),
\end{equation}
Later on we will see these corrections in the small $N$, weak environmental noise limit of other theories.

\section{Environmental stochasticity without storage mechanism} \label{maritan}

In this section we consider a similar process, now with fitness variations caused by environmental fluctuations.  In each elementary step \emph{two} individuals ($i$ and $j$) are chosen at random for a duel: with probability $p \equiv 1/2 - \eta_{i,j}/4 $ $j$ dies and is replaced by an offspring of $i$, and with probability $1-p$, $i$ dies and $j$ reproduces into the empty gap ($\eta_{i,j} = -\eta_{j,i}$). If $\eta = 0$ for all $i$ and $j$, the pure demographic game considered in the last section is recovered. $\eta_{i,j}=0$ if $i$ and $j$ are conspecific individuals,  $\eta_{i,j} =  \gamma$ if $i$ belongs to species 1  and $j$ to species 2,  $\eta_{i,j} =  -\gamma$  if $j \in 1$ and $i\in 2$.
Here we are interested in the situation of time averaged neutrality considered in \cite{kalyuzhny2015neutral,danino2016effect,hidalgo2016species}, meaning that both species have the same average fitness but $\eta$ fluctuates in time (environmental stochasticity). The simplest way to model such a scenario is to consider a \emph{dichotomous} noise: the fitness parameter  may take only two values, $\eta = -\eta_{j,i} =  \pm \gamma$ and after each elementary step the chance of the environment to stay in the same state is $1-1/\tau$, while the chance of the environment  to flip (i.e, $\pm \gamma \to \mp \gamma$) is $1/\tau$. Accordingly, the environmental fluctuations are characterized by two parameters: $\gamma$, the fluctuation amplitude and the environmental persistence time (in units of a generation) $\delta \equiv \tau/N$. Both white Gaussian noise and white Poisson noise can be recovered from the dichotomous noise by taking suitable limits \cite{ridolfi2011noise}, so the results obtained are quite generic.
For fluctuating environment we should  define again the persistence time. If, at  $t=0$ species $\#1$ is represented by $n$ individuals \emph{and} the system is in the $+\gamma$ state, the persistence time is  $T_+(n)$, while if at $t=0$ the system is in the $-\gamma$ state we denote the persistence time by $T_-(n)$.  The BKE  (with $q \equiv 1/2-\gamma/4$) reads,
\begin{eqnarray} \label{eq6}
T_+(n) = \frac{1}{N} &+& \left( 1-\frac{1}{\tau} \right) \left\lbrace F(x) \left[ q  T_+(n+1) + (1-q) T_+(n-1)\right] + [1-F(x)]T_+(n)  \right\rbrace \nonumber \\ &+& \frac{1}{\tau} \left\lbrace F(x) \left[ q T_-(n-1) + (1-q) T_-(n+1)\right] + [1-F(x)]T_-(n) \right\rbrace \\
T_-(n) = \frac{1}{N} &+& \left( 1-\frac{1}{\tau} \right) \left\lbrace F(x) \left[ q  T_-(n-1) + (1-q) T_-(n+1)\right] + [1-F(x)]T_-(n)  \right\rbrace \nonumber \\ &+& \frac{1}{\tau} \left\lbrace F(x) \left[ q T_+(n+1) + (1-q) T_+(n-1)\right] + [1-F(x)]T_+(n) \right\rbrace \nonumber
\end{eqnarray}

Defining $S(n) \equiv (T_+(n) + T_-(n))/2$,  $\Delta(n) \equiv (T_+(n) - T_-(n))/2$, moving to the continuum limit $x \equiv n/N$ and expanding $T(n \pm 1)$ to the second order in a  Taylor series as above,  one finds:
\begin{eqnarray} \label{eq7}
\frac{2}{\tau} \Delta &=& \left( 1-\frac{2}{\tau} \right) x(1-x)\left[ \frac{-\gamma}{N} S' + \frac{\Delta ''}{N^2} \right] \\ \nonumber
-1 &=& x(1-x) \left( \frac{S''}{N}  -\gamma \Delta ' \right)
\end{eqnarray}
where the derivatives are with respect to $x$.

If $\tau =2$ (i.e., $\delta = 2/N$), the chance of the environment to switch from $\pm \gamma$ to $\mp \gamma$ after each elementary step is $1/2$. In this limit the environmental stochasticity becomes demographic: the outcome of each duel is determined by two independent drawings of a random variable, one that dictates the environmental conditions and the other determines the result given these conditions, so the net chance to win a duel is again $1/2$ with no correlations in time. This can be seen from Eqs. (\ref{eq7}): the r.h.s. of the upper equation is zero, meaning that $\Delta=0$, and the lower equation reduces to (\ref{eq4}).

 In most of the realistic scenarios one would like to assure that the persistence time of the environment is independent of size of the community, i.e., to assume that $\delta$ is fixed, and so in the large $N$ limit $2/\tau$ is negligible. In this case, (\ref{eq7}) takes the form,

\begin{eqnarray}  \label{eq8}
-\gamma S' + \frac{\Delta ''}{N} &=& \frac{2\Delta}{\delta x(1-x)}   \\ \nonumber
   \frac{S''}{N}- \gamma  \Delta' &=& -\frac{1}{x(1-x)}
\end{eqnarray}

 Using a (numerically inspired) dominant balance argument we discovered that for reasonably large $N$  the $\Delta''/N$ term is negligible in the first equation (see further discussion of this point in the last paragraph of this section). Isolating  $\Delta$ from the first equation and plugging it into the second, we obtain an inhomogeneous, first order ODE for $S'$. From symmetry  $S(x)$ peaks at $x=1/2$, meaning that $S'(1/2)=0$; using that as a boundary condition one obtains:
\begin{eqnarray}  \label{eq9}
S' &=& -\frac{N ln\left(\frac{x}{1-x} \right)}{1+ N \delta \gamma^2 x(1-x)/2} \\ \nonumber
\Delta &=& \frac{N \delta \gamma x(1-x) ln\left(\frac{x}{1-x} \right)/2}{1+ N \delta \gamma^2 x(1-x)/2}.
\end{eqnarray}

$S$ is the average persistence time, when one averages over both initial conditions (i.e., starting at a certain $x$, when half of the realizations the environment favors species 1 at $t=0$) and fluctuation histories. To calculate $S$ we integrate $S'$, invoking the boundary condition $S(0)=0$ (which by symmetry implies $S(1)=0$ as well),  so that

\begin{equation}\label{eq10}
S(x) = -N \int^x_0 dt \frac{ln\left(\frac{t}{1-t} \right)}{1+ Gt(1-t)}
\end{equation}
where $$G \equiv N \delta \gamma^2/2$$ is the control parameter of the transition between the demographic  and the environmental behavior as we shall demonstrate immediately. The general form of Eq. (\ref{eq10}) is $S(x) = N {\cal F}(G)$, in agreement with the prediction of \cite{hidalgo2016species} for a one parameter scaling function. -

Eq. (\ref{eq10}) is a closed expression for $S(x)$, and one may easily evaluate this integral numerically. Moreover, an explicit solution may be written in a form of the dilogarithm functions \cite{abramowitz1964handbook}  $Li_2(x) \equiv - \int_1^x dt \  ln(t)/(t-1)$:
\begin{eqnarray}\label{eq11}
S(x) = -\frac{N}{\sqrt{G(4+G)}} && \left[2 \ \textrm{atanh}\left(\frac{\sqrt{G(4+G)}(1-x)}{2+G(1-x)}\right)\ln{(1-x)} + 2 \ \textrm{atanh}\left(\frac{\sqrt{G(4+G)}x}{2+Gx}\right)\ln{x} \right. \nonumber \\ &&- \textrm{Li}_2\left(1-\frac{2(1-x)}{G+\sqrt{G(4+G)}} \right) + \textrm{Li}_2\left(1+\left(G+\sqrt{G(G+4)}\right)(1-x)/2 \right)  \nonumber \\ &&- \textrm{Li}_2\left(1-\frac{2x}{G+\sqrt{G(4+G)}} \right) + \textrm{Li}_2\left(1+\left(G+\sqrt{G(G+4)}\right)x/2 \right) \nonumber \\ && - \left. \textrm{Li}_2\left(1+\left(G+\sqrt{G(G+4)}\right)/2 \right) + \textrm{Li}_2\left(1+\frac{G-\sqrt{G(G+4)}}{2} \right) \right].
\end{eqnarray}
To understand intuitively this result it is better to consider the asymptotic behavior of the maximum persistence time, $S(1/2)$, as obtained from  the integral (\ref{eq10}), in the limits of small and large $G$. As long as $G \ll 1$ one can neglect the second term in the denominator to get,
\begin{equation} \label{eq11a}
S(1/2) = \int_0^{1/2} -N ln\left(\frac{x}{1-x} \right) = N \ln(2),
\end{equation}
 i.e., the pure demographic result (\ref{eq5}).  If $G \gg 1$ one may define $1/G \ll \zeta \ll 1$ and use asymptotic matching \cite{bender1999advanced} to get the leading behavior:
\begin{equation}\label{eq12}
S(1/2) \approx   -N \int_0^{\zeta} \frac{ln(x)}{1+Gx} - \frac{N}{G}\int_{\zeta}^{1/2}\frac{ln\left(\frac{x}{1-x} \right)}{ x(1-x)}\approx N \left[\frac{ln^2(G)+\pi^2/3}{2G} + {\cal O} \left(\frac{ln^2(G)}{G^2}\right)\right].
\end{equation}

\begin{figure}
\includegraphics[width=16cm]{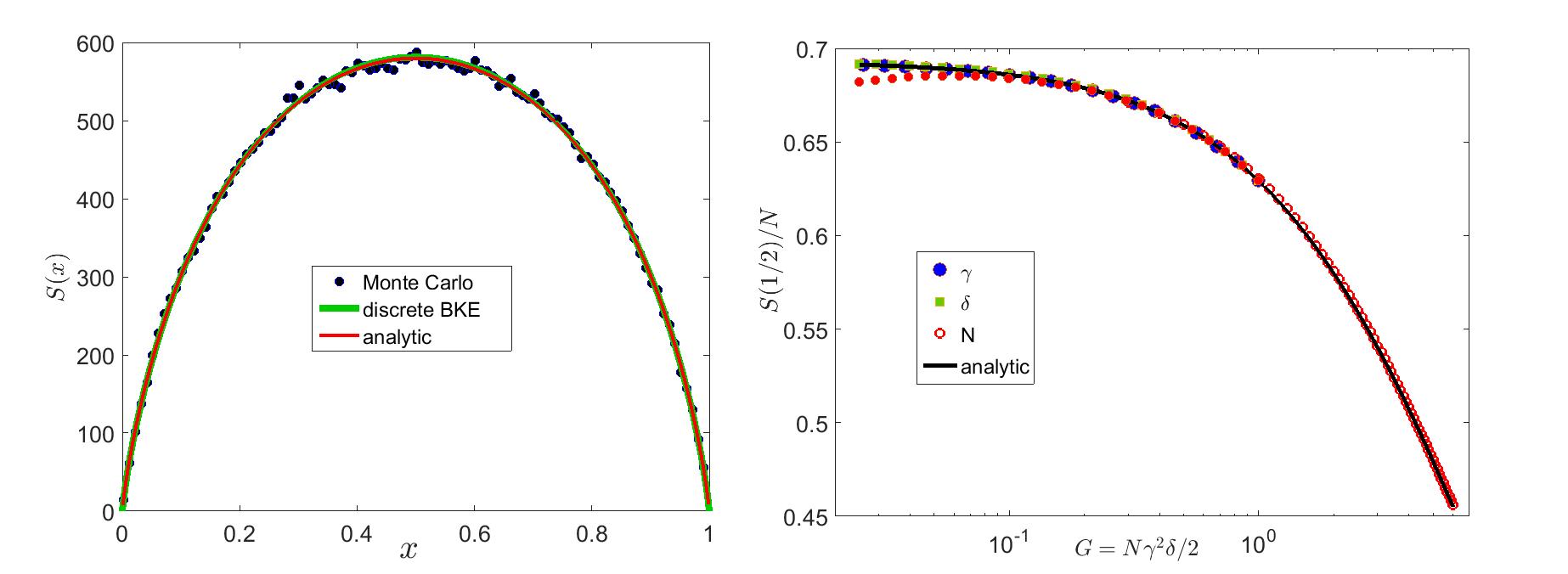}
\caption{A  comparison  between the results of a Monte-Carlo simulations, numerical solution of the BKE  and the analytic results (\ref{eq11}) for $S(x)$ is presented in the left panel. Results are depicted for a system with $N=1000$, $\gamma = 0.2$ and $\delta = 0.1$, where the MC lifetimes were averaged over $10000$ runs. In the right panel we compare the maximum lifetime of the system, $S(1/2)$, with the predictions of Eq. (\ref{eq11}), testing both the accuracy of this formula and the general scaling hypothesis suggested in \cite{hidalgo2016species}. Four sets of results, obtained using a numerical solution of the master equation, are presented. Green squares are the results for $N=2000$, $\gamma = 0.1$ and $\delta$ from $2.5 \cdot 10^{-3}$ to $0.1$. Blue circles are for $N=2000$, $\delta = 0.1$ and $\gamma$ from $0.01$ to $0.1$. Red (open and filled) circles represent results with varying $N$ with $\delta =0.1$: for the open, $\gamma= 0.2$ and $N$ runs from $70$ to $3000$, while for the filled circles $\gamma = 0.1$ and $N$ is between $50$ and $2000$. Clearly, the analytic prediction and the results are in perfect agreement as long as $N$ is sufficiently large. For small $N$ (filled red) one may identify small deviations from the theoretical prediction and a violation of the scaling hypothesis. The magnitude of these discreetness induced corrections for $N=50$ is about $0.01$, in agreement with the  ${\cal O}(1/2N)$  scaling  predicted in Eq. (\ref{eq5b}).  }\label{fig2}
\end{figure}

Figure \ref{fig2} demonstrates the validity of our results. The agreement between $S(x)$ as obtained from  Eq. (\ref{eq11}) and the numerical simulations is quite good, and the dependency of $S(1/2)/N$ on $G$  is shown to satisfy (\ref{eq11}).

As Hidalgo and coworkers \cite{hidalgo2016species} suggested the large $N$ asymptotic behavior of the persistence time scales, for this scenario,  with $\ln^2(N)$, and  the general behavior is described by a function of the form $S(1/2) = N {\cal F}(G)$, where ${\cal F}(G)$ approaches unity when the environmental noise vanishes ($G \to 0$) and  $\ln^2(G)/G$ when $G$ is large. Our expression (\ref{eq11}) provides the explicit form of the required scaling function, with the correct asymptotic behaviors (Eqs. \ref{eq11a} and \ref{eq12}).

The asymptotic matching analysis and Eq. (\ref{eq12}) allows us to identify the large $N$ scaling regime, where the $ln^2(G)$ term is much bigger than the first correction. This happens when,
\begin{equation} \label{eq12a}
N \gg \frac{2e^{\pi/\sqrt{3}}}{\delta \gamma^2} \approx \frac{12.25}{\delta \gamma^2} .
\end{equation}

Now let us consider the limitations of the general scaling analysis. As seen in Figure \ref{fig2}, when $G$ approaches zero with $N$ fixed (meaning that the environmental stochasticity parameter $\delta \gamma^2$ is decreasing) the system approaches smoothly the demographic noise limit $S(1/2)/N = \ln(2)$. However when the noise is kept fixed and $N$ decreases (filled red circles) there are deviations, since the demographic noise limit admits small $N$ corrections (\ref{eq5b}), and the scaling hypothesis breaks down. Thus, the maximum deviation is $-1/2N = -\gamma^2 \delta/4G$ which, for fixed $G$, grows with $\delta$.

Another limitation of the general scaling comes from the large $\delta$ limit. As seen in Appendix \ref{apb}, the persistence time under a  time-independent selective force (i.e., when $\eta$ stays fixed at $\eta_0$) scales as $\ln(N)/\eta_0$. This implies that, if $\delta \gg \ln(N)/|\gamma|$, absorption may take place during a single sweep of the environment. Indeed our numerics shows that, in this regime, the predictions of this section are violated and one should stick to the $\ln(N)/|\gamma|$ estimation.

A surprising outcome of (\ref{eq11}) is the persistence time of a single mutant, $S(1/N)$. Evaluating  (\ref{eq11}) for $x = 1/N$ and approximating it for large $N$ we get,
\begin{equation} \label{eq11c}
S(1/N) =\frac{\ln(1+\gamma^2 \delta/2)}{\gamma^2 \delta/2} \ln(N).
\end{equation}
 This result implies that in the large $N$ limit, as long as the noise amplitude $\gamma^2 \delta/2$ is small, the clonal interference threshold is \emph{the same} for this model and for a model with pure demographic stochasticity. One may understand this result intuitively by looking at the process as a random walk in the log-abundance space \cite{kessler2015neutral}. The first passage probability at $t$  scales like $t^{-3/2}$ with a cutoff at $\ln^2(N)$, so the average lifetime, starting at one, is $\ln(N)$.

Finally, we would like to discuss the dominant balance argument that allows us to neglect the $\Delta''/N$ term in the first equation of (\ref{eq8}). As seen from Eqs. (\ref{eq9}) and (\ref{eq10}), any derivative with respect to $x$ contributes a factor of $G$ to an otherwise ${\cal O}(1)$ expression (this may be seen easily by defining a new parameter, $y \equiv Gx$ and considering small values of $x$, but the argument is valid in general). Accordingly, as long as $\gamma^2 \delta^2 \ll 1$, the $\Delta''/N$ term is negligible with respect to the other two terms.

\section{Environmental stochasticity with storage effect} \label{chesson}

As discovered by Chesson and collaborators~\cite{chesson1981environmental,hatfield1989diffusion,hatfield1997multispecies}, the introduction of  environmental noise into the system may induce an attractive force that stabilizes the system. Even if the environmental fluctuations are ``neutral" (i.e., the fitness of both species, when averaged over time, is the same), they may support the invasion (and recovery) of low abundance species. Accordingly, one expects that the persistence time of the system is large compared to the ${\cal O}(N)$ dependence of the pure demographic noise case. In this section we consider this scenario for a two-species community with dichotomous noise.

The conditions under which this ``storage effect" takes place, and its dependence on the system parameters, were considered by us in \cite{danino2016effect}. Here we would like to find the persistence time obtained from this dynamics. We have chosen the dynamics of the Chesson-Warner ``lottery game" \cite{chesson1981environmental}. These authors have analyzed the problem without demographic stochasticity, so extinction and fixation were not allowed, while here the game is analyzed with demographic noise.

For the sake of concreteness, let us consider a forest with $N$ trees that belong to two different species. Every tree produces the same number of seeds, such that the density of seedlings (of a given species) per unit area reflects the relative abundance of the species'  trees in the forest. During each elementary step, one tree is chosen at random to die, and the seedlings  compete to fill the vacant site. The environmental conditions determine the fitness of these seedlings: if there are $n$ trees of species $1$ and $N-n$ trees of species $2$, the chance of species $1$ to capture an empty slot, $P_1$, will be:
\begin{equation} \label{eq15}
P_1 = \frac{n}{n+e^{\eta}(N-n)}=\frac{x}{x+e^{\eta}(1-x)},
\end{equation}
and the chance of species $2$ will be $1-P_1$.

$\eta$ measures the relative fitness of species $1$: if $\eta=0$ the chance of a species to fill the gap is proportional to its relative abundance in the forest, and  the game reduces to two species dynamics with pure demographic stochasticity. When $\eta<0$ species $1$ is preferred as its per seedling chance to take over the gap is larger, while if $\eta>0$ this chance is smaller. Again we are interested in the case where $\eta = \pm \gamma$, where $\gamma$ measures the strength of the environmental fluctuations. As in the last section, the chance to switch between plus and minus $\gamma$ is $1/\tau$ per elementary step, where $\delta \equiv \tau/N$ is the environment persistence time as measured in units of a generation time.

Defining $P_1^+$ and $P_1^-$ as the values of $P_1$ in the $\pm \gamma$ states, correspondingly, one realizes that the chance of species $1$ to increase its abundance during an elementary step is $(1-x)P_1^+$ when the environment favors it and $(1-x)P_1^-$ when the fitness of species $2$ is higher. The backward Kolomogorov equation then reads,
\begin{eqnarray} \label{eq16}
T_+(n) = \frac{1}{N} &+& \left( 1-\frac{1}{\tau} \right) \left\lbrace  \left[ (1-x)P_1^+  T_+(n+1) + x (1-P_1^+)T_+(n-1)\right] + [1-(1-x)P_1^+ -x (1-P_1^+) ]T_+(n)  \right\rbrace \nonumber \\ &+& \frac{1}{\tau} \left\lbrace  \left[ (1-x)P_1^-  T_-(n+1) + x (1-P_1^-)T_-(n-1)\right] + [1-(1-x)P_1^- -x (1-P_1^-) ]T_-(n) \right\rbrace \\
T_-(n) = \frac{1}{N} &+& \left( 1-\frac{1}{\tau} \right) \left\lbrace  \left[ (1-x)P_1^-  T_-(n+1) + x (1-P_1^-)T_-(n-1)\right] + [1-(1-x)P_1^- -x (1-P_1^-) ]T_-(n)  \right\rbrace \nonumber \\ &+& \frac{1}{\tau} \left\lbrace  \left[ (1-x)P_1^+  T_+(n+1) + x (1-P_1^+)T_+(n-1)\right] + [1-(1-x)P_1^+ -x (1-P_1^+) ]T_+(n) \right\rbrace. \nonumber
\end{eqnarray}

Defining $S(n) = (T_+(n) + T_-(n))/2$, $\Delta(n) = (T_+(n) - T_-(n))/2$, moving to the continuum limit $x \equiv n/N$,  expanding $T(n \pm 1)$ in a Taylor series as above and expanding to second order in $\gamma$   one finds:
\begin{eqnarray} \label{eq17}
\frac{2}{\delta x(1-x)} \Delta &=& \left( 1-\frac{2}{N\delta} \right) \left\lbrace \gamma \left( \frac{1}{2}-x \right)\frac{S''}{N}+ \gamma S' + \left[ 1+ \gamma^2 \left( \frac{1}{2}-x \right)^2 \right] \frac{\Delta''}{N} + \gamma^2 \left( \frac{1}{2}-x \right) \Delta' \right\rbrace \\ \nonumber
-\frac{1}{x(1-x)} &=& \gamma \left( \frac{1}{2}-x \right)\frac{\Delta''}{N}+ \gamma \Delta' + \left[ 1+ \gamma^2 \left( \frac{1}{2}-x \right)^2 \right] \frac{S''}{N} + \gamma^2 \left( \frac{1}{2}-x \right) S'.
\end{eqnarray}

Eq. (\ref{eq17}) may be discussed in two different limits. The first is the case where $\tau$ is ${\cal O}(1)$, i.e., when $\delta \sim 1/N$. Unlike the case without storage considered in section \ref{maritan}, where  under fast switching of the environment the noise becomes essentially demographic, here in this limit the storage effect is very strong and the lifetime of the system, as we shall see, is exponential in $N$ when $N \to \infty$. The other, more relevant case, is when $\delta$ is fixed as $N$ grows (meaning that the persistence time of the environmental fluctuations is independent of the size of the community); in this limit the storage effect is weaker, and the persistence time scales like a positive power of $N$ in the asymptotic limit.

\subsection{Strong storage effect}

To consider a strong effect case, let us assume $\tau = 2$. This implies that after an elementary timestep the environment switches with probability $1/2$, so there is effectively  no  persistence of the environmental conditions. In this scenario $\delta = 2/N$, meaning that the r.h.s. of the first equation in (\ref{eq17}) vanishes, so $\Delta =0$ ($\Delta$ measures the difference between $T^+(x)$ and $T^-(x)$, and there is no such difference if the environmental conditions are uncorrelated). Eqs. (\ref{eq17}) then reduce to
\begin{equation}\label{eq18}
S''+ N \gamma^2 \left( \frac{1}{2}-x \right) S'=-\frac{N}{x(1-x)},
\end{equation}
where the $\gamma^2$ term in the coefficient of $S''$ was neglected with respect to unity, since we assumed that $\gamma$ is small. Eq. (\ref{eq18}) is an inhomogeneous, first order equation for  $W \equiv S'$; using an integrating factor to solve it, plugging the boundary condition $W(1/2)=0$ (meaning that the maximum lifetime of the system occurs at $x=1/2$ since the two species are symmetric) and integrating again one finds,
\begin{equation}\label{eq20q}
S(x) = -N\int_0^{x}dt e^{-N \gamma^2 t(1-t)/2} \int_t^{1/2} dq \frac{e^{N \gamma^2 q(1-q)/2}}{q(1-q)},
\end{equation}
where the boundary condition $S(0)=0$ determines the lower bound of the outer integral.

As long as the quantity $g \equiv N \gamma^2/2$ is large, the main  contribution to the inner integral comes from the peak close to $1/2$ (the $1/q$ divergence when $q$ approaches zero is regularized by the outer integral). The Laplace method then yields, for the inner integral in the large $g$ limit,
$$ \int_t^{1/2} dq \frac{e^{N \gamma^2 q(1-q)/2}}{q(1-q)} \sim 2e^{g/4} \sqrt{\frac{\pi}{g-4}}.$$

\begin{figure}
\includegraphics[width=7cm]{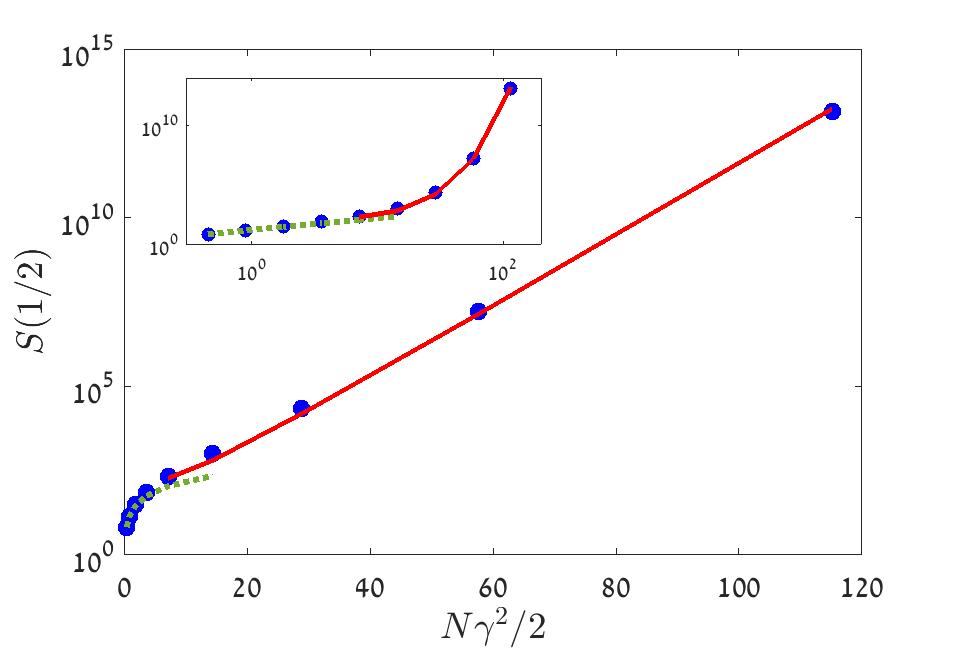}
\caption{Maximum persistence time, $S(1/2)$, as a function of $g \equiv N\gamma^2/2$, for different values of $N$ ($N = 2^n \cdot  10$ for $n=1..9$) and $\gamma=0.3$. Blue circles are  the result obtained from a numerical solution of the BKE, red line is the large $g$ asymptotic expression (\ref{eq20z}) and dashed green line shows the $N \log(2)$, demographic noise, relationship.  The inset shows the same data on a double logarithmic scale, emphasizing the breakdown of (\ref{eq20z}) and the demographic noise behavior at small values of $g$.}
\label{fig3}
\end{figure}

The outer integration in the large $g$ limit is  trivial, since only the low $t$ region contributes and one may replace $t(1-t)$ by $t$ and extend the limits of integration to infinity. Accordingly,
\begin{equation}\label{eq20z}
S(1/2) \sim \frac{1}{\gamma^2} \sqrt{ \frac{8 \pi}{N \gamma^2 - 8}} e^{N \gamma^2/8}.
\end{equation}
As suggested by the denominator of the square root, this expression is valid only for $g \gg 4$. In the small $g$ regime  the logarithmic divergence close to zero dominates the inner integral and the problem converges to its demographic noise limit, as demonstrates in Figure \ref{fig3}.

\subsection{Weak Storage effect}

Now let us consider the case where $\delta$ is fixed while $N \to \infty$, so that the $2/(N\delta)$ in the upper line of  Eq. (\ref{eq17}) is negligible, meaning that we are dealing with the more realistic situation where $\delta$, the correlation time of the environment, is fixed in units of a generation and  is independent of the community size.

As in the case without the storage effect, it turns out that only the $S'$ and the $\Delta$ term are important in the first equation of (\ref{eq17}), and the $\Delta''$ is negligible in both equations. This numerical observation may be justified, as before, as long as $\gamma^2 \delta$ and $\gamma^2 \delta^2$ are much smaller than one.  Accordingly, Eqs.  (\ref{eq17})  reduce to,
\begin{eqnarray} \label{eq19}
\frac{2}{\delta x(1-x)} \Delta &=&  \gamma S'  \\ \nonumber
-\frac{1}{x(1-x)} = \gamma \Delta' &+& \frac{S''}{N} + \gamma^2 \left( \frac{1}{2}-x \right) S',
\end{eqnarray}
where we also omitted the $\gamma^2$ term in the coefficient of $S''$, since it is small compared to one.

\begin{figure}
\includegraphics[width=16cm]{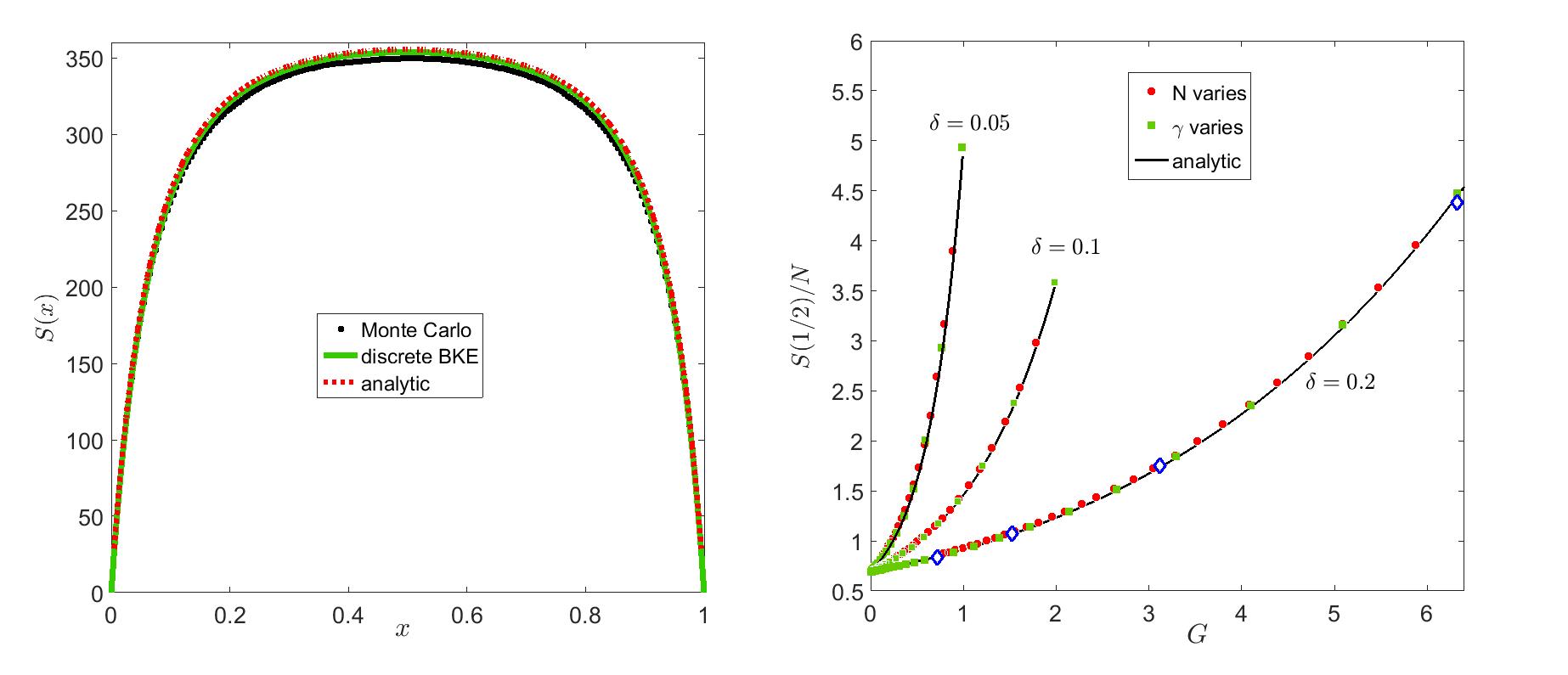}
\caption{In the left panel, a  comparison between the results of a Monte-Carlo simulations, numerical solution of the BKE  and the analytic results (i.e., numerical evaluation of \ref{eq23}) for $S(x)$, is presented. Results are shown for $N=200$, $\gamma = 0.4$ and $\delta = 0.2$. The MC lifetimes were averaged over $250000$ runs. In the right panel we compare the maximum lifetime of the system, $S(1/2)$, with the predictions of Eq. (\ref{eq23}), testing, again, both the accuracy of this formula and the general scaling hypothesis. Three full black lines, corresponding to the outcomes of (\ref{eq23})  for different values of $\delta$ (appearing next to each line), are presented along with the results obtained using numerical solutions of the BKE. For the two sets $\delta = 0.05$ and $\delta =0.1$, the green squares are the numerical results with $N=1000$ and $\gamma$ varies between $0.01$ and $0.2$, and the red circles represent the results for $\gamma = 0.2$ and $N \in [50..1000]$. For $\delta = 0.2$,  green squares correspond to $N=400$ and $\gamma \in [0.01..0.4]$ while the red circles are $\gamma = 0.4$ and $N \in [50..400]$. All lines converge to $ln(2)$, the demographic noise limit, as $G$ goes to zero. Open diamonds are the lifetimes obtained from Monte-Carlo simulation of the process for $\delta =0.2 \ \gamma =0.4 \ N \in [50,100,200,400]$, where each realization was averaged over at least $100000$ runs.}\label{fig100}
\end{figure}

The upper line of (\ref{eq19}) implies $\gamma \Delta'(x) = \gamma^2 \delta \left[ x(1-x)S''/2+(1/2-x)S'\right]$. Plugging that into the second equation one finds,
\begin{equation} \label{eq20}
\gamma^2 N (1+\delta) \left( \frac{1}{2}-x \right) S' +  \left(1+\frac{N \gamma^2 \delta x(1-x)}{2}\right)S''=-\frac{N}{x(1-x)},
\end{equation}
or, with $W \equiv S'$ and $F_1(x) \equiv 1+N \gamma^2 \delta x(1-x)/2$,
\begin{equation} \label{eq21}
W'(x)F_1(x)+W(x)F_1'(x)\frac{1+\delta}{\delta}=-\frac{N}{x(1-x)}.
\end{equation}
Multiplying both sides of (\ref{eq21}) by the integrating factor $F_1^{1/\delta}$ one may write (\ref{eq21}) as
 \begin{equation} \label{eq21b}
\frac{d}{dx} \left(W(x)F_1^{1+1/\delta}(x)\right)=-N\frac{F_1^{1/\delta}(x)}{x(1-x)}.
\end{equation}
One integration over $x$ yields $W$, where (by symmetry)  the limits of integration are such that $W(1/2)=0$. Accordingly,
 \begin{eqnarray} \label{eq23}
S(x) &=& N \int_0^x dt F_1^{-1-1/\delta}(t)\int_t^{1/2} dq \frac{F_1^{1/\delta}(q)}{q(1-q)} \\ \nonumber &=& N \int_0^x dt \left(1+Gt(1-t)\right)^{-1-1/\delta}\int_t^{1/2} dq \frac{\left(1+Gq(1-q)\right)^{1/\delta}}{q(1-q)},
\end{eqnarray}
with $G \equiv N \gamma^2 \delta/2$. Figure \ref{fig100} shows the agreement between (\ref{eq23}) and the numerics.

Let us try to extract the large $N$ asymptotic behavior of the maximum persistence time $S(1/2)$. In the inner integral of (\ref{eq23}) the integrand is growing exponentially with $q$ (for small $q$ the numerator grows like $exp(Gq/\delta)$) so the main contribution  comes from the regime $q>\delta/G$. As before, although the denominator contribution diverges logarithmically when the lower limit of the integral approaches zero (reflecting the effect of demographic noise), this divergence is $G$-independent and is regularized by the outer integral. This observation allows us to factor the numerator of the inner integral,
\begin{eqnarray}
\int_t^{1/2} dq \frac{\left(1+Gq(1-q)\right)^{1/\delta}}{q(1-q)} &=& G^{1/\delta} \int_t^{1/2} dq \left[q(1-q) \right]^{-1+1/\delta} \left(1+\frac{1}{Gq(1-q)} \right)^{1/\delta} \\ \nonumber &\sim& G^{1/\delta} \left(1+\frac{4}{G} \right)^{1/\delta} \int_t^{1/2} dq [q(1-q)]^{-1+1/\delta}.
\end{eqnarray}
where in the last step we replaced the $1/q(1-q)$ term in the last factor  by its value at $q=1/2$, the region from which the main contribution comes (the same result may be obtained by Laplace integration around the maximum point at $q=1/2$). Accordingly,
\begin{equation} \label{eq26}
S(x) = N G^{1/\delta} \left(1+\frac{4}{G} \right)^{1/\delta} \int_0^x  dt \frac{1}{ \left[1+Gt(1-t)\right]^{1+1/\delta}} \left[B_{\frac{1}{2}}(\frac{1}{\delta},\frac{1}{\delta})-B_{t}(\frac{1}{\delta},\frac{1}{\delta})\right].
\end{equation}
Here $B_z(a,b)$ is the incomplete Beta function.

The main contribution to the outer integral (\ref{eq26}) is  from the $t \to 0$ regime, so $1-t \approx 1$. In this limit $B_{t}(\frac{1}{\delta},\frac{1}{\delta}) \approx t^{1/\delta}$. Accordingly, the maximum persistence time as $N \to \infty$  is,
\begin{equation} \label{eq27}
S(1/2) \sim N B_{\frac{1}{2}}(\frac{1}{\delta},\frac{1}{\delta}) \left(1+\frac{4}{G} \right)^{1/\delta}  \delta G^{1/\delta-1} = \frac{2}{\gamma^2} B_{\frac{1}{2}}(\frac{1}{\delta},\frac{1}{\delta}) \left(1+\frac{8}{\gamma^2 \delta N} \right)^{1/\delta} \left(\frac{\gamma^2 \delta}{2} N\right)^{1/\delta}.
\end{equation}
  This expression yields a very good approximation to the exact expression (\ref{eq23}) as long as $G$ is large. In particular, the term $(1+4/G)^{1/\delta}$ behaves in the large $G$ limit like $exp(4/\delta G)$ and converges to one, yielding the asymptotic power-law behavior $S(1/2) \sim N^{1/\delta}$ predicted by \cite{hidalgo2016species}. One sees, however, that this asymptotic emerges only when $G \delta = N \gamma^2 \delta^2/2 \gg 4$.

\begin{figure}
\includegraphics[width=10cm]{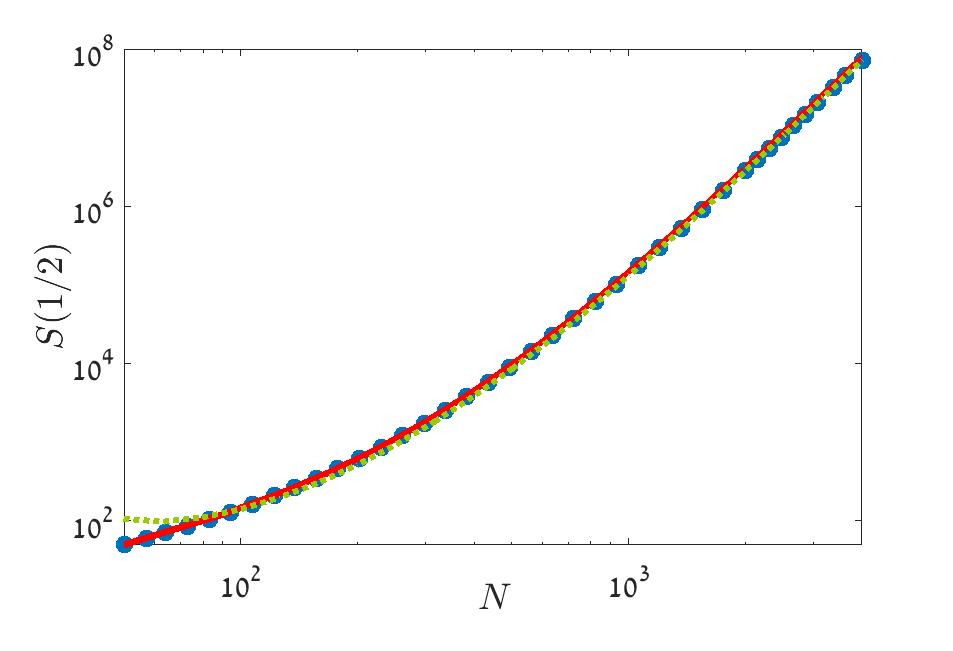}
\caption{Persistence time for a system with weak storage effect ($\delta =0.2, \ \gamma = 0.5$) is shown against $N \in [50,...,4000]$ on a double logarithmic scale. Blue circles are numerical results obtained from the BKE, Red full line was obtained from numerical integration of   (\ref{eq23}) and the green dashed line is the approximation (\ref{eq27}). As predicted, the slope approaches $1/\delta=5$ in the asymptotic limit, but to reach this limit $4/G\delta$ should be smaller than one. This condition is fulfilled here around $N=800$. The approximation (\ref{eq27a}) holds for a much wider regime. }\label{fig4}
\end{figure}

  When a single mutant is introduced into the system, the persistence time is obtained from the integral (\ref{eq26}) with an upper limit at $x = 1/N$. The result is
   \begin{equation}\label{eq27a}
   S(1/N) = \frac{\gamma^2}{2} S(1/2) =B_{\frac{1}{2}}(\frac{1}{\delta},\frac{1}{\delta}) \left(1+\frac{8}{\gamma^2 \delta N} \right)^{1/\delta} \left(\frac{\gamma^2 \delta}{2} N\right)^{1/\delta},
   \end{equation}
   and one can see, now, how the $N^{1/\delta}$ asymptotics suggested by \cite{hidalgo2016species} emerges. The approximation suggested in Eq. (\ref{eq27a}) captures a much wider range, as demonstrated in Fig. \ref{fig4}.

  Unlike the case with no storage, now the persistence time of a single mutant has the same $N^{1/\delta}$ scaling as the maximum persistence time. Accordingly, the threshold to clonal interference occurs at much smaller values of $\nu$, of order $N^{-(1+1/\delta)}$.

\section{Storage effect against selection} \label{selection}

For a system exposed to external migration, or allowing for a constant rate of mutations or speciation events, a strictly neutral/symmetric model, where the species fitnesses are assumed to be exactly equal (either literally or after  averaging over time) is quite implausible. One should expect that phenotypic differences lead to some value of time-independent selective advantage of one of the species. Even mechanisms like emergent neutrality \cite{kessler2014neutral,vergnon2012emergent} yield competitive communities with slight selective variations between species.

When there is no storage in the system, selection determines the large $N$  limit of the persistence time, and the effect of environmental stochasticity becomes irrelevant. Without storage, stochasticity yields only a random walk in the log-abundance space and constant, directional bias wins against such a random movement in the long run. In Appendix \ref{apc} we present an analysis of this case: the model considered in section \ref{maritan}, superimposed on fixed selective bias, and the large $N$ result is indeed identical to the result of a model with selection and without environmental stochasticity, analyzed in Appendix \ref{apb}.

A much more interesting question emerges when the storage effect, which gives stability to the system, interferes with a fixed selective advantage of one species. This question has been considered by Chesson and Warner \cite{chesson1981environmental} who concluded that their conditions for coexistence is fulfilled (the probability distribution function is normalizable)  if the stochasticity is strong with respect to the strength of selection and satisfies: $\gamma^2 > 2 \eta_0$. Here we consider the same problem from the persistence time perspective; we will show that the stability condition presented by \cite{chesson1981environmental} is, in fact, equivalent to the requirement that the persistence time scales superlinearly with $N$.

 To model this scenario we change the dynamics of $\eta$ in Eq. (\ref{eq15}): instead of jumping from $+\gamma$ to $-\gamma$, $\eta$ jumps between $\eta_0+\gamma$ and $\eta_0-\gamma$, meaning that species 2 has a fixed selective advantage $\eta_0$, superimposed on the environmental fluctuations of amplitude $\gamma$.

Plugging this definition of $\eta$ and $P_1$ into Eq. (\ref{eq16}), one finds, to the leading order in $\eta_0$,
\begin{eqnarray} \label{eq40}
\frac{2}{\delta x(1-x)} \Delta &=& \left( 1-\frac{2}{N\delta} \right) \left\lbrace \gamma \left( \frac{1}{2}-x \right)\frac{S''}{N}+ \gamma S' + \left[ 1+ \gamma^2 \left( \frac{1}{2}-x \right)^2 \right] \frac{\Delta''}{N} + \gamma^2 \left( \frac{1}{2}-x \right) \Delta' \right\rbrace \\ \nonumber
-\frac{1}{x(1-x)} &=& \gamma \left( \frac{1}{2}-x \right)\frac{\Delta''}{N}+ \gamma \Delta' + \left[ 1+ \gamma^2 \left( \frac{1}{2}-x \right)^2 \right] \frac{S''}{N} +\left[ \gamma^2 \left( \frac{1}{2}-x \right) - \eta_0 \right] S'.
\end{eqnarray}
Note that $\eta_0$ appears only in the last term of the lower equation. As before, dominant balance analysis allows us to simplify these equations,
\begin{eqnarray} \label{eq41}
\frac{2}{\delta x(1-x)} \Delta &=&  \gamma S'  \\ \nonumber
-\frac{1}{x(1-x)} = \gamma \Delta' &+& \frac{S''}{N} + \left[ \gamma^2 \left( \frac{1}{2}-x \right) - \eta_0 \right] S'.
\end{eqnarray}
  With the definitions $W \equiv S'$, $F_1(x) \equiv 1+G x(1-x)$, $G\equiv N\gamma^2\delta/2$,
\begin{equation} \label{eq42}
W'(x)F_1(x)+W(x)\left(-\eta_0 N +F_1'(x)\frac{1+\delta}{\delta} \right)=-\frac{N}{x(1-x)}.
\end{equation}
This is a first order equation for $W$ that may be solved with an integration factor, however since the problem is no longer symmetric the point at which $W$ vanishes is not at $x = 1/2$. Labeling this point $x^*$, the result for $S$, taking into account the boundary condition $S(0) =0$, is,
\begin{eqnarray} \label{eq43}
S(x)  = N \int_0^x dt \frac{e^{-z(t)}}{\left(1+Gt(1-t)\right)^{1+1/\delta}}\int_t^{x^*} dq \frac{e^{z(q)}\left(1+Gq(1-q)\right)^{1/\delta}}{q(1-q)},
\end{eqnarray}
with
\begin{equation}
z(x) \equiv -\mu \  atanh \left((2x-1)\sqrt{\frac{G}{G+4}}\right),
\end{equation}
where
\begin{equation}
\mu = \frac{2\eta_0 N}{\sqrt{G(G+4)}} \approx \frac{4\eta_0 }{\gamma^2 \delta},
\end{equation}
the last approximation holding for large $G$.

\begin{figure}
\includegraphics[width=10cm]{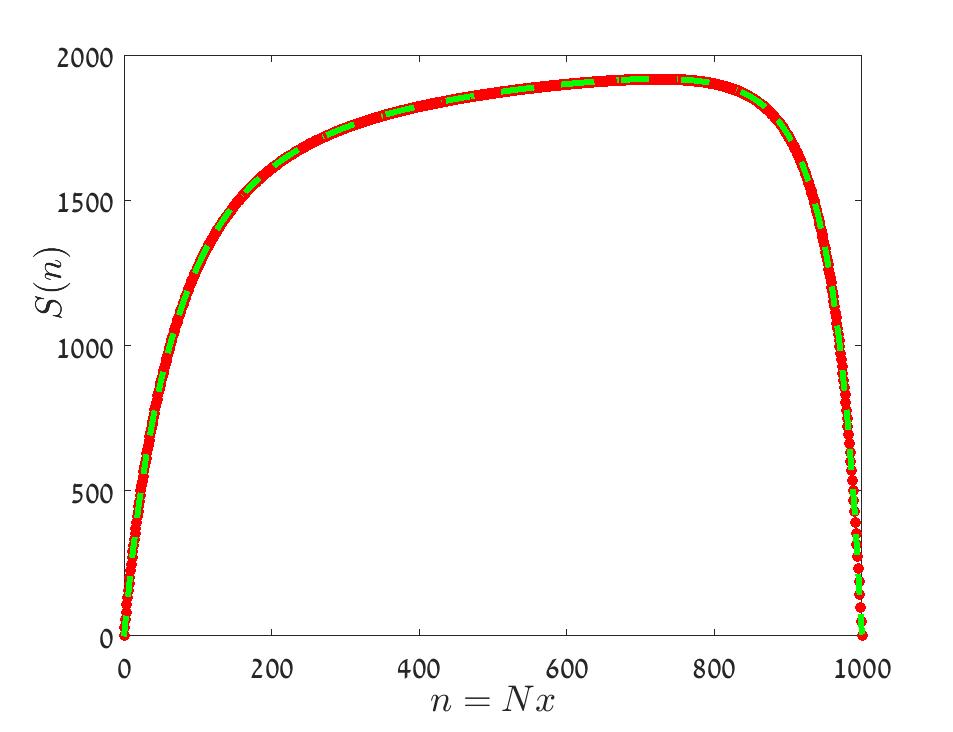}
\caption{Persistence time for a system with storage effect ($\delta =0.1, \ \gamma = 0.2, \ N=1000$) and selection $\eta_0 = 5 \cdot 10^{-3}$. The result obtained from numerical solution of the master equation (filled red circles) and those obtained from a numerical integration  of (\ref{eq43}) with $G=2$ (dashed green line) are shown. The value  $x^*=0.723$ was obtained numerically from the condition $S(1)=0$. As explained in the text, positive $\eta_0$ gives an average selective advantage to species 2, hence the lifetime of the system peaked to the right of $x=1/2$. }\label{fig10}
\end{figure}

To proceed, we first need to determine $x^*$, using the  second boundary condition, $S(1)=0$. Solving numerically for $x^*$,  $S(x)$ may be plotted against the numerical solutions of the BKE  and the fit is very nice (Fig. \ref{fig10}).

To obtain the large $N$ behavior of $S(x^*)$ from (\ref{eq43}), we assume, as in the former section, that the main contribution of the inner integral comes from finite values of $q$ (again, although the integrand blows like $1/q$ at zero, this contribution is only logarithmic and is regularized by the outer integration). Expressing the hyperbolic tangent in terms of logarithms, the inner integral may be written as,
\begin{eqnarray} \label{eq44a}
\int_t^{x^*} dq \frac{e^{z(q)}\left(1+Gq(1-q)\right)^{1/\delta}}{q(1-q)} &=& \\ \nonumber G^{1/\delta} \int_t^{x^*} &dq& \frac{\left[q(1-q)\right]^{1/\delta}}{q(1-q)} \left(\frac{q}{1-q}\right)^{-\mu/2}\left(1+\frac{1}{Gq(1-q)}\right)^{1/\delta}
\left(\frac{\left(1+(2q-1)\sqrt{\frac{G}{G+4}}\right)(1-q)}
{\left(1-(2q-1)\sqrt{\frac{G}{G+4}}\right)q}\right)^{-\mu/2},
\end{eqnarray}
Notice that the last two terms under the integral approach unity in the large $G$ limit. Accordingly, we use for the Laplace integration of (\ref{eq44a}) only the terms that are important when $G$ is large,
\begin{eqnarray} \label{eq45}
 \int_0^{x^*} dq  \ q^\alpha (1-q)^\beta.
\end{eqnarray}
The maximum  of the integrand in (\ref{eq45}) occurs at  $ \overline{q} = \alpha/(\alpha + \beta)$, where  $\alpha = -1+1/\delta - \mu/2$ and $\beta = -1+1/\delta + \mu/2$. Note that, for positive values of $\mu$, the maximum of $S$, $x^*$, moves to the right of $1/2$ (see figure \ref{fig10}) while the maximum of the integrand occurs at $\overline{q}<1/2$. This implies that (unless $\mu$ is very small, see below)  the peak is within the range of integration.

Plugging $q= \overline{q}$ in the last two terms and using Laplace integration one obtains for the inner integral,
\begin{eqnarray} \label{eq46}
\left(1+\frac{1}{G\overline{q}(1-\overline{q})}\right)^{1/\delta}
\left(\frac{\left(1+(2\overline{q}-1)\sqrt{\frac{G}{G+4}}\right)(1-\overline{q})}
{\left(1-(2\overline{q}-1)\sqrt{\frac{G}{G+4}}\right)\overline{q}}\right)^{-\mu/2} \sqrt{\frac{2 \pi \alpha \beta}{(\alpha+\beta)^3}} \  \overline{q}^\alpha(1-\overline{q})^\beta G^{1/\delta}.
\end{eqnarray}

  To deal with the outer integral, one notices that the contribution of the outer integral comes mainly from the $t\ll 1/G$ region. In that case,  $z(t)$ should be expanded for small $t$ and large $G$, such that  $Gt$ is fixed, yielding
\begin{equation}
z(t) \approx  -\frac{\mu}{2} \ln(x+1/G).
 \end{equation}
Since the inner integral is now a constant, the contribution of the outer integral will be
\begin{equation}
 \int_0^{x^*} dt \frac{e^{-z(t)}}{\left(1+Gt(1-t)\right)^{1+1/\delta}} \approx  G^{-\frac{\mu}{2}}\int_0^\infty \frac{dt}{  \left(1+Gt\right)^{-1+\mu/2 - 1/\delta}} \approx   \frac{G^{-\frac{\mu}{2}}}{G(1+1/\delta -\mu/2)}.
 \end{equation}

The final form of the large $N$ approximation is (the absolute value signs will soon be justified),
\begin{equation}\label{eq47}
S(x^*) \sim \left(1+\frac{1}{G\overline{q}(1-\overline{q})}\right)^{1/\delta}
\left(\frac{\left(1+(2\overline{q}-1)\sqrt{\frac{G}{G+4}}\right)(1-\overline{q})}
{\left(1-(2\overline{q}-1)\sqrt{\frac{G}{G+4}}\right)\overline{q}}\right)^{-\mu/2} \sqrt{\frac{2 \pi \alpha \beta}{(\alpha+\beta)^3}} \  \frac{2\overline{q}^\alpha(1-\overline{q})^\beta}{\gamma^2 \delta (1+1/\delta -|\mu|/2)} G^{\frac{1-|\tilde{s}|}{\delta}}.
\end{equation}
Here we define the dimensionless selection parameter mentioned above,
$$\tilde{s} \equiv \frac{2 \eta_0}{\gamma^2},$$
this parameter measuring the strength of the selection in units of the environmental stochasticity.

\begin{figure}
\includegraphics[width=10cm]{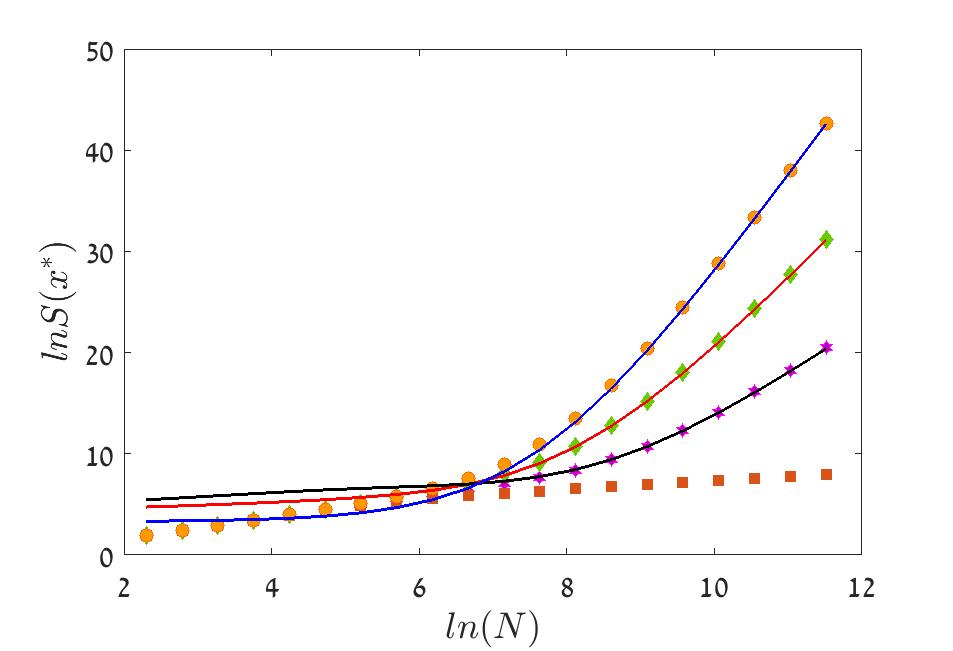}
\caption{Log of the maximum persistence time, $S(x^*)$, is plotted versus $\ln(N)$ for a system with storage effect ($\delta =0.1, \ \gamma = 0.2$) and selection. Orange circles correspond to  $\eta_0 = \tilde{s}= 0$, green diamonds represent $\eta_0 = 0.005, \  \tilde{s}= 1/4$, purple hexagrams are  for $\eta_0 = 0.01, \  \tilde{s}= 1/2$ and the brown squares are $\eta_0 = 0.02, \  \tilde{s}= 1$. The data obtained from numerical integration of Eq. (\ref{eq43}) and the full lines are the large $N$ approximation (\ref{eq47}) for each case.      }\label{fig11}
\end{figure}

The terms in the first two parentheses approach unity at large $G$, but are important when $G$ is small. When all other parameters are kept fixed and $N$ increases the lifetime of the community growth like $N^{\frac{1-|\tilde{s}|}{\delta}}$. All these features are demonstrated in Figure \ref{fig11}. As explained above, the Chesson-Warner condition $\tilde{s}<1$ is translated to a condition for superlinear growth of $S$ with $N$ in the large $N$ limit.

Comparing (\ref{eq47}) with (\ref{eq27}), one notices that the main effect of selection in the large $N$ limit is to decrease the exponent of $N$, from $1/\delta$ to $(1-|\tilde{s}|)/\delta$. When $\tilde{s}=1$ there is no asymptotic power-law behavior, as demonstrated in  Fig. \ref{fig11}. When $\tilde{s} \to \infty$ the $N$ scaling has to be logarithmic (see Appendix \ref{apb}).

Until now we have assumed that the values of $\eta_0$, $\mu$ and $\tilde{s}$ are all positive, but in Eq. (\ref{eq47}) we take the absolute values of these quantities in the last terms. The reason for that is as follows. When $\eta_0$ is negative $x^*$ occurs to the left of $1/2$, while $\overline{q}>1/2$. The inner integration may be done using Laplace integrals only if the peak  is inside the integration region, i.e., it should include the region between $x^*$ and $x=1$, meaning that the outer integral has to be done over the range between  $x$ and $1$.   The main contribution to the outer integral in this range comes from values of $t$ around one, so the sign of $\mu$ in the expression for $z(t)$  is inverted. Accordingly, the quantities that depend on $\mu$ from the outer integral appear with an absolute value sign. The $\mu$ in the second parentheses of  (\ref{eq47}) comes from the inner integration and do not change sign, but since $\overline{q}(\eta_0) = 1- \overline{q}(-\eta_0)$, the result is the same.

The discussion that leads to Eq. (\ref{eq47}) was based on the assumption that the inner integration region contains most of the Gaussian peak around $\overline{q}$. This is not exactly true when $1/\mu$ and $\delta$  are not small, as both $\overline{q}$ and $x^*$ approach $1/2$, and in the extreme limit $\mu=0$ only half of the peak is integrated, meaning that (\ref{eq47}) should be multiplied by $1/2$. In general one would like to multiply (\ref{eq47}) by $(1+\textrm{Erfc}\left[(x^*-\overline{q})/\sigma\right])/2$, where $\sigma = \sqrt{\alpha \beta/(\alpha+\beta)^3}$.

\section{Discussion}

Coexistence of many competing species in a local community and the maintenance of multiple alleles in a gene pool are ubiquitous in natural systems. Simple explanations to these phenomena, like strong niche differentiation (in ecology) or heterozygote advantage (in population genetics), are not plausible in many of the systems considered so far. The search for alternative coexistence mechanisms  became a subject of intensive research.

The (temporal) storage effect, suggested by Chesson and coworkers, is apparently an appealing candidate.  Environmental stochasticity is almost always quite strong in biosystems (see, e.g. \cite{hekstra2012contingency}, where environmental stochasticity dominates system fluctuations even under extremely stable external conditions. See also \cite{kalyuzhny2014niche,kalyuzhny2014temporal,chisholm2014temporal} for an analysis of ``standard" high diversity ecosystems, showing that the main driver is environmental noise). Therefore, the fact that it may become a stabilizing factor, supporting the growth of rare species and providing an effective frequency dependent selective mechanism is very appealing.

From a different perspective, the ubiquitous presence of environmental stochasticity leads to attempts to incorporate it into the neutral model - one of the main theoretical frameworks in  both population genetics and community ecology \cite{chisholm2014temporal,kalyuzhny2014niche,kalyuzhny2015neutral,danino2016effect}. Biodiversity under neutral dynamics relays on speciation-absorption equilibrium, and as we
have seen here, environmental stochasticity affects strongly the absorption rates.
 The works published so far have relied on a mix of numerical experiments and a few analytic arguments.  A general theoretical understanding of the time averaged neutral model, comparable with  classical results that were obtained for the neutral model with demographic stochasticity  (like the Fisher log-series and zero-sum multinomials),  is still missing.

Herein we have extended the results of \cite{hidalgo2016species} and discussed a few models that incorporate environmental stochasticity, selection and demographic noise, for a two species community. These results are of interest in and of themselves, as they provide a new insight to the classical works of \cite{chesson1981environmental,hatfield1989diffusion,hatfield1997multispecies}, shedding light on the question of a community's persistence time. Moreover, our results provide an answer to a fundamental question: at what rate one should introduce new types to the system in order to maintain it in a diverse state.

There are still many open questions, including the behavior of higher moments of the persistence time for the model with storage (without storage, or with pure demographic noise, the variance of the persistence time is equal to the time itself, as one can understand from the Galton-Watson theory or from the theory of first passage time, correspondingly). The problem of  the fixation time (which is assumed to determine the pace of the evolutionary process) is also of interest.

However, we believe that the most important aspect of the work presented here is as a first step towards an analytic theory of neutral dynamics with environmental stochasticity.  Generalization of our work  from the two species case to a many species community will allow one to find the species abundance distributions and the species richness for the time averaged neutral theory of biodiversity in stable mutation-extinction equilibrium, and even to incorporate weak effects of selection into the otherwise symmetric model (as we did numerically in \cite{danino2016effect}). We intend to address these topics in subsequent publications.

{\bf Acknowledgments} N.M.S. acknowledge the support of the Israel
Science Foundation, grant no. $1427/15$. D.A.K. acknowledge the support of BSF grant no.  2015619.

\appendix

\section{Methods} \label{methods}

Through this work we compare the analytic results, obtained from the backward Kolomogorov equations (BKE), to the result obtained from three types of numerics: agent-based Monte-Carlo simulation of the birth-death process,  numerical solutions of the BKE itself and numerical integrations of expressions like (\ref{eq43}).

\subsection{Monte-Carlo simulations}

The agent based simulations were performed as follows. We start with a community of $N$ individuals, in which  $xN$  belong to species $\#1$ and $N(1-x)$ belong to species $\#2$.

To simulate a neutral dynamics with pure demographic stochasticity (Section \ref{demo}), two individuals are chosen at random in each elementary step, with probability $1/2$ the first dies and is replaced by an offspring of the second (meaning that after the replacement both individuals belong to the species of the second) and with probability $1/2$ the second dies and is replaced by the offspring of the first. After each elementary step the time is incremented by $1/N$.  The persistence time of the community, or the persistence time $T(x)$, is defined as the mean time elapsed until one of the species goes extinct, and we measured it by running many histories with the same initial condition and averaging over the outcomes.

For the system with fixed selection, or with environmental noise with no storage effect (Appendix \ref{apb}, Section \ref{maritan}), the same procedure has been used, where now if two individuals belonging to different species are involved in a duel, the chance of each to win is determined by its relative fitness. The relative fitness $\eta$  is time independent in fixed environment and fluctuates between the values $\pm \gamma$ in the environmental noise case.

The storage effect is considered in Section \ref{chesson} and the interplay between  the storage effect and  the presence of overall selective advantage is considered in Section \ref{selection}. In both cases the dynamics is slightly different: in each elementary step one individual is chosen at random to die, and then the species that is recruited to fill the gap is determined according to the abundance weighted by the fitness, as set out in Eqs. (\ref{eq15}).

\subsection{Numerical solution of the backward Kolomogorov equation}

The discrete BKEs considered along this paper, like Eqs.  (\ref{eq6})  and (\ref{eq16}), are second order, linear, inhomogeneous difference equations that have the general form,
$$-\frac{1}{N} \left[ \begin{array}{c} 1\\ \vdots \\ \vdots \\ \vdots \\ \vdots \\ 1 \end{array} \right] = \begin{bmatrix} W_{1^+,1^+} &  W_{1^+,2^+} & \cdots & W_{1^+,1^-} & W_{1^+,2^-} & \cdots \\  \vdots & &\ddots & &\vdots \\  \vdots & &\ddots & &\vdots\\ W_{1^-,1^-} &  W_{1^-,2^-} & \cdots & W_{1^-,1^+} & W_{1^-,2^+} & \cdots \\  \vdots & &\ddots & &\vdots \\  \vdots & &\ddots & &\vdots  \end{bmatrix} \times \left[ \begin{array}{c} T^+_1 \\ \vdots \\ T^+_{N-1} \\ T^-_1 \\ \vdots \\ T^-_{N-1}   \end{array} \right],$$
where $W_{n,m}$ is the rate of transitions from the $n$ to the $m$ state. Accordingly, the values of $T^{\pm}_n$ [and consequently the values of $S(n)$ and $\Delta(n)$] may be determined by inverting this $2N-2 \times 2N-2$ matrix and multiplying the outcome by the constant vector $-1/N$.

\section{Pure selection  without environmental stochasticity} \label{apb}

In this appendix we would like to calculate the persistence time of a mutant species of fitness $\eta_0$ with $n$ individuals in a two species community of size $N$ under fixed selective pressure (no environmental fluctuations) with demographic stochasticity.

The model is the same model considered in section \ref{maritan} for environmental stochasticity with no storage effect; each elementary step two individuals are chosen at random for a duel. If they belong to the same species nothing happens, if they belong to different species then species $\#1$ wins with probability $q$ and species $\#2$ wins with  probability $1-q$. The loser dies and an offspring of the winner takes its slot, so $n \to n \pm 1$. The environment is fixed, meaning that $q = 1/2 -\eta_0$. When $\eta_0 \to 0$, the result should converge to (\ref{eq5}), while if the effect of selection is strong one expects the asymptotic $log(N)/\eta_0$ dependency.

The BKE may be derived from (\ref{eq6}) by taking $\tau \to \infty$  (as before $F=2x(1-x)$ where $x=n/N$)

\begin{eqnarray} \label{eq51}
T(n) = \frac{1}{N} +F(x) \left[ q  T(n+1) + (1-q) T(n-1)\right] + [1-F(x)]T(n)
\end{eqnarray}
Expanding $T$ in a Taylor series and keeping derivative up to the second order one obtains,
\begin{equation} \label{eq51a}
 T'' + \eta_0 N T' = -\frac{N}{x(1-x)},
\end{equation}
The solution to this equation is immediate. Using an integrating factor we get $$\left(T'\exp(\eta_0 N x)\right)' = -N\exp(\eta_0 N x)/(x(1-x)),$$ and this yields,
\begin{equation} \label{eq99}
T(x) = N \int_0^x dt e^{-\eta_0 N t} \int_t^{x^*} dq \frac{\exp(\eta_0 N q)}{q(1-q)},
\end{equation}
where $x^*$ is determined by the condition $T(1) = 0$ (see section \ref{selection}). If $\eta_0 =0$, $x^* = 1/2$ and the integrals yield the demographic noise result  $ T(x) =  -N[x\log(x)+(1-x)\log(1-x)]$, as expected.

The integrals (\ref{eq99}) are doable even for finite $\eta_0$, yielding a result in terms of hyperbolic arctangents and exponential integrals. To find $x^*$ using the condition $T(1) = 0$ we used this analytic solution, plugged in the limits at zero and one, expanded the result  for large $\eta_0 N$ and solved for $x^*$ to find,
\begin{equation}
 x^* \sim \frac{\ln\ln(\eta_0 N)}{\eta_0 N}.
 \end{equation}
 Since $x^*$ is small in the large $N$ limit, the $(1-q)$ term in the denominator of the inner integral may be neglected if one looks for the maximum persistence time. The leading term for this quantity is,
 \begin{equation} \label{eq55a}
  T(x^*) \sim  \frac{2 \ln(\eta_0 N)}{\eta_0 }.
 \end{equation}

\section{Selection and environmental stochasticity without storage} \label{apc}

Here we consider the effect of a time independent selective advantage of species $\#1$, when superimposed on environmental stochasticity, without storage. Practically we are looking at the model considered in section \ref{maritan}, when $\eta$ jumps between $\eta_0 + \gamma$ and $\eta_0 - \gamma$, so $\eta_0$ is the time independent component of the fitness.  Since we have no storage the dynamics resembles that of a random walker (in log abundance space) with fixed bias towards one of the edges, so when $N$ goes to infinity one expects  that $S_{max} \sim ln(N)/\eta_0$, while for small $N$s and weak $\eta_0$ the $ln^2(N)$ behavior found in section \ref{maritan} dominates.

The BKE  (with $q_1 \equiv 1/2+\gamma/4+\eta_0/4$ ($q_1$ is the chance to jump to the right in the plus phase) and $q_2 \equiv 1/2+\gamma/4-\eta_0/4$) ($q_2$ is the chance to jump to the left in the minus phase)  reads,
\begin{eqnarray} \label{eq60}
T_+(n) = \frac{1}{N} &+& \left( 1-\frac{1}{\tau} \right) \left\lbrace F(x) \left[ q_1  T_+(n+1) + (1-q_1) T_+(n-1)\right] + [1-F(x)]T_+(n)  \right\rbrace \nonumber \\ &+& \frac{1}{\tau} \left\lbrace F(x) \left[ q_2 T_-(n-1) + (1-q_2) T_-(n+1)\right] + [1-F(x)]T_-(n) \right\rbrace \\
T_-(n) = \frac{1}{N} &+& \left( 1-\frac{1}{\tau} \right) \left\lbrace F(x) \left[ q_2  T_-(n-1) + (1-q_2) T_-(n+1)\right] + [1-F(x)]T_-(n)  \right\rbrace \nonumber \\ &+& \frac{1}{\tau} \left\lbrace F(x) \left[ q_1 T_+(n+1) + (1-q_1) T_+(n-1)\right] + [1-F(x)]T_+(n) \right\rbrace \nonumber
\end{eqnarray}

\begin{figure}
\includegraphics[width=10cm]{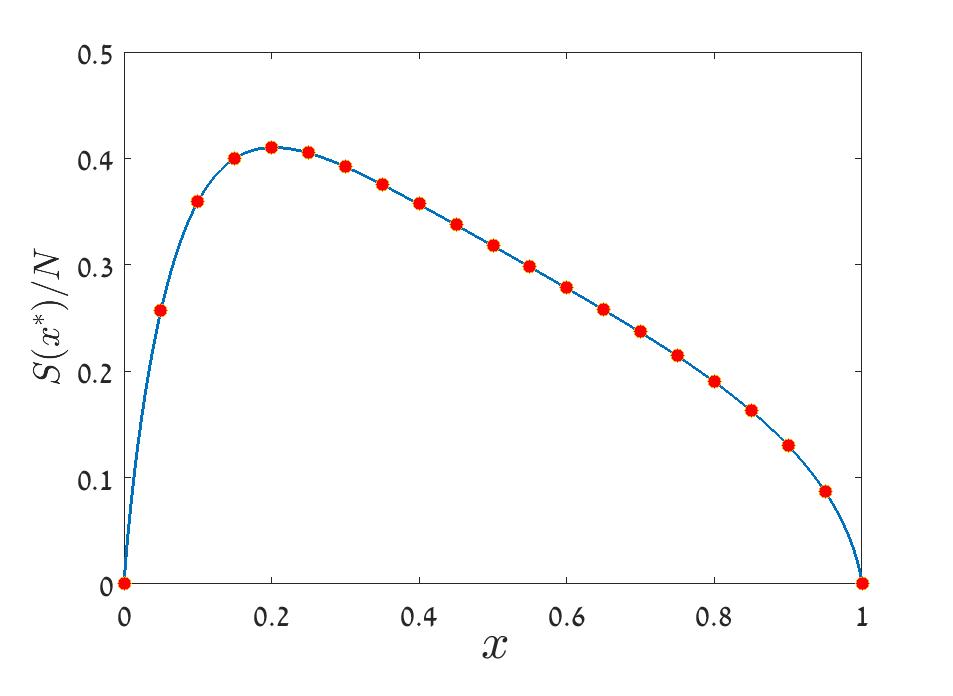}
\caption{ Maximum persistence time, $S(x^*)/N$, is plotted versus $x$ for a system with no storage effect ($\delta =0.1, \ \gamma = 0.1, \ N = 1000$) and selection ($\eta_0 = 0.01$). Red circles were obtained from numerical solution of the integrals (\ref{eq64}), where  the value of $x^*$ was found from the condition $S(1)=0$. The full blue line was obtained from the numerical solution of the BKE with the same parameters.}\label{fig64}
\end{figure}

Defining $S(n) = (T_+(n) + T_-(n))/2$, $\Delta(n) = (T_+(n) - T_-(n))/2$, moving to the continuum limit and expanding $T(n \pm 1)$ to the second order in a  Taylor series as above,  one finds, after  substituting $x \equiv n/N$:
\begin{eqnarray} \label{eq61}
\frac{2 \Delta }{\tau x(1-x)} &=& \left( 1-\frac{2}{\tau} \right)\left[ \frac{\gamma}{N} S' + \frac{\Delta ''}{N^2} + \eta_0 \frac{\Delta'}{N} \right] \\ \nonumber
-\frac{1}{x(1-x)} &=&  \frac{S''}{N}  +\gamma \Delta' + \eta_0 S'.
\end{eqnarray}
For small $\delta$ this yields (see section \ref{maritan}),
\begin{eqnarray} \label{eq62}
\frac{2 \Delta}{\delta x(1-x)}  &=&  \gamma S' + \frac{\Delta''}{N} + \eta_0 \Delta'\\ \nonumber
-\frac{1}{x(1-x)} &=&  \frac{S''}{N}  +\gamma \Delta' + \eta_0 S'.
\end{eqnarray}
Neglecting the $\Delta''$ and the $\Delta'$ terms in the upper equation, solving for $\Delta$ in terms of $S'$, $\Delta = \gamma \delta x(1-x) S'/2$, and plugging the expression for $\Delta'$ into the lower equation, we obtained,
\begin{equation} \label{eq63}
-\frac{N}{x(1-x)} = [1+Gx(1-x)]S'' + [\eta_0 N + G (1-2x)]S'.
\end{equation}
Solving this equation using an integrating factor, the expression for $S$ is,
\begin{equation} \label{eq64}
S(x) = N \int_0^x dt \frac{\left(\frac{1-2t-\sqrt{1+4/G}}{1-2t+\sqrt{1+4/G}}\right)^{-\mu/2}}{1+Gt(1-t)}\int_t^{x^*} dq \frac{\left(\frac{1-2q-\sqrt{1+4/G}}{1-2q+\sqrt{1+4/G}}\right)^{\mu/2}}{q(1-q)},
\end{equation}
where $G = N \gamma^2 \delta/2$ is the same parameter used in section \ref{maritan}, $x^*$ is the value of $x$ where $S$ reaches its maximum and
\begin{equation}
\mu \equiv \frac{2 \eta_0 N}{\sqrt{G(G+4)}}.
\end{equation}
When $G \to 0$ Eq. (\ref{eq63}) converges to (\ref{eq51a}) for a system with demographic noise and selection but without environmental stochasticity, while if $\eta_0 \to 0$ we recover  Eq. (\ref{eq8}). $\mu$ reflects the relative strength of  selection in comparison with the environmental noise; for large $G$, $$\mu \to 4\eta_0/(\gamma^2 \delta).$$
Figure \ref{fig64} shows the fit between the numerical solution of the BKE and the results obtained from a numerical solution of the integral \ref{eq64}.

To illustrate how selection dominates the behavior when $N \to \infty$, one note that in this limit $\sqrt{(4+G)/G} \sim 1+2/G$,  so (\ref{eq64}) may be written as,
\begin{equation} \label{eq66}
S(x) = N \int_0^x dt \frac{\left(\frac{1-t}{1+Gt}\right)^{\mu/2}}{1+Gt(1-t)}\int_t^{x^*} dq \frac{(1+Gq)^{\mu/2}}{q(1-q)^{1+\mu/2}},
\end{equation}
 Let us assume, for the moment, that while $x^*$ approaches zero when $N$ is large, $Gx^* \gg 1$ in this limit. In such a case the $1+Gq$ may be replaced by $Gq$ in the numerator of the inner integral of  (\ref{eq66}), since the main contribution is from the upper bound. The inner integral is then solvable, and the contribution from the outer integral may be evaluated by  calculating it in three different regimes: for $0<t<\zeta$ ($\zeta \ll 1, \ G \zeta \gg 1$), $1-\epsilon <t<1$ (again ($\epsilon  \ll 1, \ G \epsilon \gg 1$) and $\zeta <t < 1-\epsilon$ snd using asymptotic matching. To satisfy $S(1) = 0$ one finds that $x^*$ must fulfil,
 \begin{equation}
 \left(\frac{x^*}{1-x^*}\right)^{\mu/2} \frac{1}{\mu G} = \frac{\ln(G)}{G^{1+\mu/2}},
 \end{equation}
 i.e. (assuming, again, $x^*  \ll 1$),
 $$ (x^*)^{\mu/2} \sim \mu \frac{\ln(G)}{G^{\mu/2}}.$$
 Note that the assumption $Gx^* \gg 1$ turns out to be self consistent. Evaluating the outer integral of (\ref{eq66}) from zero to $x^*$ yields the community persistence time as $N \to \infty$,
 \begin{equation} \label{eq70}
 S(x^*) \sim \frac{2\ln(\eta_0 N)}{\eta_0}.
 \end{equation}
as obtained in Appendix \ref{apb} for selection without demographic stochasticity [Eq. (\ref{eq55a})].

\end{document}